\documentclass[acmtog]{acmart}

\acmPrice{15.00}
\setcopyright{acmlicensed}
\acmJournal{TOG}
\acmYear{2017}
\acmVolume{36}
\acmNumber{4}
\acmArticle{1}
\acmMonth{7}
\acmDOI{http://dx.doi.org/10.1145/8888888.7777777}

\setcitestyle{square}

\acmConference[SIGGRAPH Asia'18]{ACM SIGGRAPH conference}{Tokyo} 

\usepackage{float}
\usepackage{booktabs} 
\usepackage{enumerate}
\usepackage{graphicx}
\usepackage{subcaption}
\usepackage{mathrsfs} 
\usepackage{wrapfig}

\def\ie{{\it i.e.,\ }}

\def\eg{{\it e.g.,\ }}

\newcommand{\mypara}[1]{\vspace{0.02in} \noindent \textbf{#1}}

\begin{document}

\title[Interactive Garments]{Learning a Shared Shape Space for Multimodal Garment Design}

\author{Tuanfeng Y. Wang} 
\affiliation{\institution{University College London}}
\author{Duygu Ceylan} 
\affiliation{\institution{Adobe Research}}
\author{Jovan Popovic} 
\affiliation{\institution{Adobe Research}}
\author{Niloy J. Mitra} 
\affiliation{\institution{University College London}}

\begin{abstract}
Designing real and virtual garments is becoming extremely demanding with rapidly changing fashion trends and increasing need for synthesizing realistic dressed digital humans for various applications. This necessitates creating simple and effective workflows to facilitate authoring sewing patterns customized to garment and target body shapes to achieve desired looks.  
Traditional workflow involves a trial-and-error procedure wherein a mannequin is draped to judge the resultant folds and the sewing pattern iteratively adjusted until the desired look is achieved. This requires time and experience. 
Instead, we present a data-driven approach wherein the user directly indicates desired fold patterns simply by sketching while our system estimates corresponding garment and body shape parameters at interactive rates. The recovered parameters can then be further edited and the updated draped garment previewed. Technically, we achieve this via a novel shared shape space that allows the user to seamlessly specify desired characteristics across multimodal input {\em without} requiring to run garment simulation at design time. 
We evaluate our approach qualitatively via a user study and quantitatively against test datasets, and demonstrate how our system can generate a rich quality of on-body garments targeted for a range of body shapes while achieving desired fold characteristics. 
\end{abstract}


\begin{teaserfigure}
  \includegraphics[width=\textwidth]{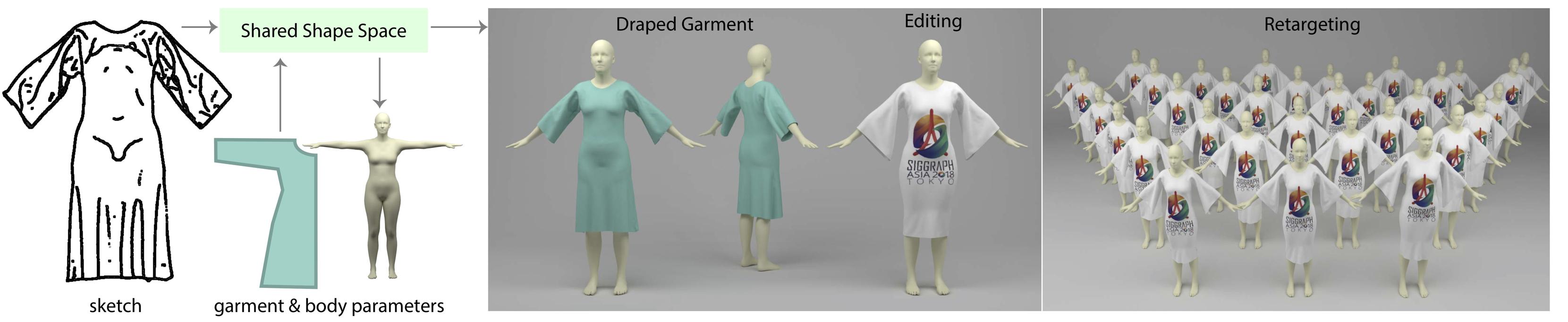} 
  \caption{Garment design is complex, given the requirement to satisfy multiple specifications including target sketched fold patterns, 3D body shape, or 2D sewing patterns and/or textures. We learn a novel {\em shared shape space} spanning different input modalities that allows the designer to seamlessly work across the multiple domains to interactively design and edit garments \textit{without} requiring access to expensive physical simulations at design time, and retarget the design to a range of body shapes.}
  \label{fig:teaser}
\end{teaserfigure}

\maketitle

\section{Introduction}

Developing effective tools for designing both real and virtual garments is becoming increasingly crucial. In today's digital age, consumers are a single-click away from online clothing stores, with an increasing appetite for new fashion styles. Similarly, virtual garment design attracts increasing interest from the entertainment industry since it is a significant component of creating realistic virtual humans for movies, games, and VR/AR applications. Both of these trends are creating a demand for fast, effective, and simple tools to design, edit, and adapt garments to various body shapes. 

Traditionally, designing real and virtual garments has been a complex, iterative, and time-consuming process consisting of multiple steps. First, the designer sketches the look and feel of a garment or alternatively drapes fabric on a physical dress form. Then a professional pattern maker creates the 2D garment patterns referred to as {\em sewing patterns}. A sample garment is made from the sewing patterns and tested by draping on a real or simulating on a virtual mannequin. Often the garment parameters need to be iteratively adjusted followed by redraping or resimulation until the desired look is achieved. Finally, in order to ensure an operational outfit, the mannequin is animated to see how the garment drapes across various body poses. Furthermore, the same style garment needs to be adapted for different body proportions through a process called \emph{pattern grading}. This essentially requires the complex and iterative process of garment design to be repeated multiple times.

\begin{figure}[t!]
  \includegraphics[width=\columnwidth]{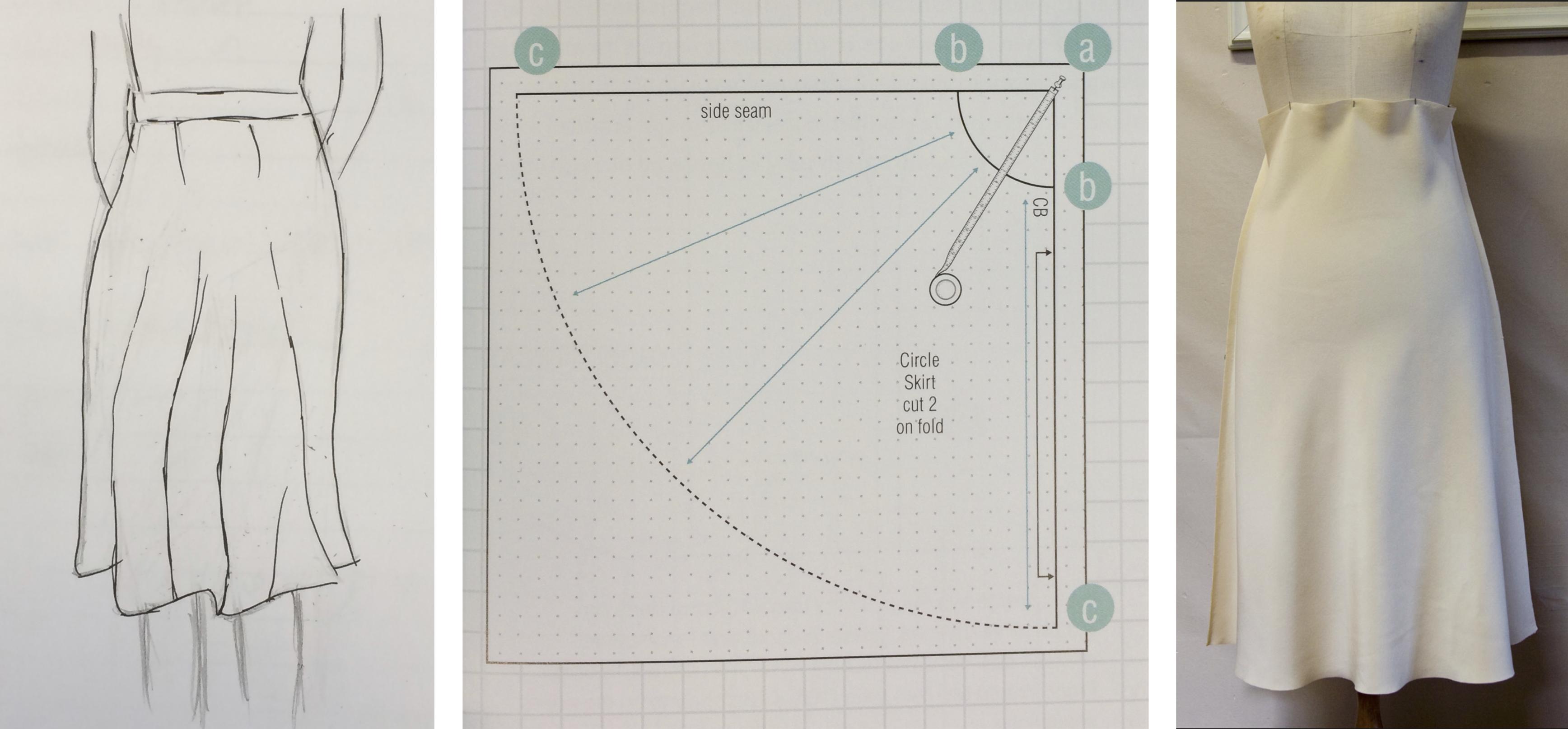}
  \caption{Garment designers often practice with different modalaties including sketch, parameter domain (e.g., parameters of 2D sewing patterns), and draped garments in 3D.}
  \label{fig:motivation}
\end{figure}

Garment design is a complex process mainly due to the fact that it operates across three different spaces, namely 2D sketches for initial design, 2D sewing patterns and material selection for parameterized modeling, and 3D garment shape to model the interaction between garments and subject bodies producing {\em on-body garments} (see Figure~\ref{fig:motivation}). 
Much of pattern making involves using various rule-based specialized cuts and stitches (e.g., darts, pleats, yokes) to achieve desired folds on the final draped garment. Note that a particularly challenging scenario is designing free flowing garments where the characteristic patterns arise due to the interaction between fabric {\em hidden} behind the creases and boundary conditions induced by the underlying body shape. To aid with such challenging scenarios, an ideal computational tool should allow the designer to freely navigate across the three design spaces and effectively capture the interaction between them.
%

Designing such a unified treatment of the three spaces has remained elusive due to several challenges. First, interpreting 2D garment sketches requires a good understanding of what shapes and folds are possible in 3D. Although the problem appears to be badly ill-conditioned, as humans, we regularly use our experience of typical garment folds and looks to `regularize' the problem and interpret artist sketches. Second, the relation between the 2D garment parameters and the final 3D shape of the garment is highly non-linear depending not only on the shape of the garment itself but also its material properties, the pose, and the shape of the 3D body it is being draped on. This necessitates a computationally heavy cloth simulation process to visualize patterns arising out of garment folds, creases, and ruffles. Finally,  modeling in 3D needs to ensure the physical plausibility of the garment by mapping the 3D designs to (near-) developable 2D sewing patterns.

We present a data-driven approach that overcomes the above challenges by learning a \textit{shared latent space} that, for the first time, unifies 2D sketches; parameters of 2D sewing patterns, garment materials, and 3D body shapes; and the final draped 3D garments. We achieve this by \textit{jointly} training multiple encoder-decoder networks that each specializes at linking pairs of representations (e.g., recovering garment and body parameters from a sketch or recovering the on-body shapes of garments from the parameters) while operating at a common embedding. To train these network, we create a large scale synthetic dataset. Specifically, we first define a set of parameterized garment types (\texttt{shirt}, \texttt{skirt}, and \texttt{kimono}) and generate different garments by sampling this representation. Then, we simulate each garment on a set of 3D body shapes and poses sampled from a deformable body model. Finally, for each of these simulated examples, we generate 2D sketches using non-photorealistic rendering. Thus, each examplar triplet in our dataset includes (i)~a 2D sketch, (ii)~garment and body parameters, and (iii)~the resultant draped 3D shape of the garment. Subsequently, by jointly training multiple encoder-decoder networks via a novel multimodal loss function, we learn a common embedding that can be queried using any of the different modalities. 

The learned shared latent space enables several  applications by linking the different design spaces. For example, starting from an input 2D sketch, we can (i)~automatically infer garment and body shape parameters; (ii)~predict the resultant 3D garment shapes from these parameters {\em without} going through an expensive cloth simulation process; (iii)~directly texture the final garments using the linked 2D sewing pattern parameterization; (iv)~sample from or interpolate in the latent space to generate  plausible garment variations; or (v) retarget 2D garment patterns to new body shapes such that the resultant on-shape garments retain the original fold characteristics. 
%
At any stage of the design and retargeting process, the garment and body parameters inferred by our method can be provided to a cloth simulator to generate the physically accurate shape of the garment on the 3D body. Figure~\ref{fig:teaser} shows several examples.

We qualitatively and quantitatively evaluate our approach against groundtruth test data, and demonstrate our interactive garment sketching for various applications. 
We also provide comparisons with alternative methods and show favorable performance~(see Section~\ref{sec:results}).
In summary, our main contributions are: 
(i)~a method that learns a joint embedding of different garment design spaces; 
(ii)~inferring garment and body parameters from single sketches;
(iii)~estimating the 3D draped configurations of the garments from the garment and body parameters to enable an interactive editing workflow; and 
(iv)~facilitating fold-aware pattern grading across different body shapes via a novel garment retargeting optimization.

\section{Related Work}

\noindent\textbf{Garment modeling.} Traditional garment design is a complex process. Starting with initial 2D sketches of the desired garment, the designer creates the necessary flat sewing patterns which are then stitched together into garment. Physical draping or physics-based simulation is used to infer the final shape of the garment. While there exist many professional tools (e.g., Optitex~\cite{optitext}, Marvelous Designer~\cite{marvelous}) to assist designers with the  process, there has been a significant effort in the research community to propose alternative computational tools.

\begin{figure*}[t!]
  \includegraphics[width=\textwidth]{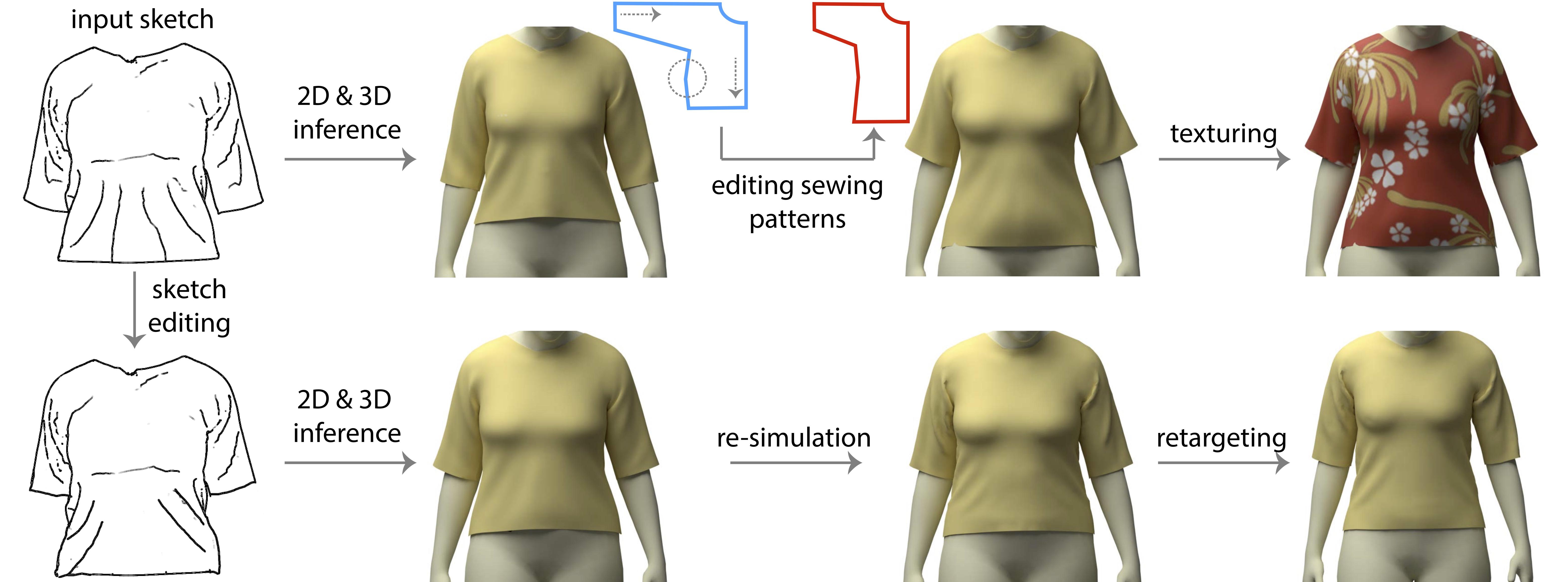}
  \caption{Given an input sketch, our network infers both the 2D garment sewing patterns (in blue)  and the draped 3D garment mesh together with the underlying body shape. Edits in the 2D sewing patterns (e.g., shorter sleeves, longer shirt as shown in red) or the sketch (bottom row) are interactively mapped to updated 2D and 3D parameters. The 3D draped garment inferred by our network   naturally comes with uv coordinates, and thus can be directly textured. The network predictions can be passed to a cloth simulator to generate the final garment geometry with fine details. Finally, the designed garment can be easily retargeted to different body shapes while preserving the original style (i.e., fold patterns, silhouette) of the design.}
  \label{fig:workflow}
\end{figure*}

Many of the previous works have specifically focused on bridging the gap across two different design spaces. For example, while Berthouzoz et al.~\shortcite{Berthouzoz:2013} focus on automatically parsing 2D garment patterns to 3D, other methods have focused on modeling 3D garments via input sketches. Several sketch-based interfaces~\cite{Turquin:2004,decaudin:2006,Robson2011} have been proposed where silhouette and fold edges in an input sketch are analyzed to create a 3D garment. However, they assume the input sketch is provided with respect to a given 3D mannequin and use the body shape of the mannequin to lift the sketch to 3D. The recent approach of Jung et al.~\shortcite{Jung:2015} provides a more general tool for modeling 3D developable surfaces with designed folds but require multi-view sketch input (e.g., frontal, side, and optionally top). The freeform surface modeling tool, BendSketch~\cite{Li:2017}, is capable of modeling plausible 3D garments from user sketches but has no notion of the corresponding 2D sewing patterns. Recently, Li et al.~\shortcite{Li:2018} present a modeling system where the user can sketch different types of strokes on an existing 3D draped garment to denote different types of desired folds. The system then extends and optimizes the 2D garment patterns such that the draped garment exhibits the desired folds. While this system provides impressive results, it assumes an initial 3D draped garment to be provided and it requires certain experience for the users to learn and place different stroke types. In contrast, our system automatically infers the parameters of the garment and the body from an input sketch and maps them to the final 3D draped shape of the garment. We achieve this by learning a joint embedding that, for the first time, unifies all three design spaces, i.e., 2D sketches, garment and body parameters, and the 3D draped shape of a garment. Subsequently, the method of Li et al.~\shortcite{Li:2018} can be used to refine cuts on the 2D garment pattern produced by our approach.

In another line of work, researchers have proposed to use other types of input such as images or RGB-D scans to model garments. Given an input image, Zhou et al.~\shortcite{Zhou:2013} first estimate 3D body shape in a semi-automatic fashion and then lift extracted silhouette edges of the garments to 3D. Jeong et al.~\shortcite{Jeong:2015} use an image of a garment on top of a mannequin as input and detect silhouette edges and landmark points to create a 3D garment. Chen et al.~\shortcite{Chen:2015} model garments from RGB-D data as a combination of 3D template components based on a set of rules. Given a database of garment templates, Yang et al.~\shortcite{Yang:2016} propose a relatively complex pipeline to determine the parameters of these garments to match an input image. More recently, Dan\v{e}\v{r}ek et al.~\shortcite{DGS:2017} present a deep learning based method which predicts the deformation of a garment from a reference 3D garment given an input image. 
While the proposed approaches provide plausible garments, they often fail to capture the details, i.e., the exact fold patterns observed in the input. In contrast, our goal is to be able to reconstruct such folds to enable realistic design and editing of garments.

\noindent\textbf{Garment editing and retargeting.} In addition to modeling garments from scratch, several methods have been proposed for editing the shape and appearance of them. Umetani et al.~\shortcite{Umetani:2011} propose an interactive editing system that enables bi-directional editing between 2D garment patterns and 3D draped forms. Bartle et al.~\shortcite{Bartle:2016} present a method to map 3D edits to a garment to plausible 2D garment patterns. In contrast, we support a multi-modal design and editing paradigm, specifically focusing on modeling the desired folds and silhouettes of a garment.

Retargeting the style of an existing garment to body shapes with different proportions is a specific form of editing that has  attracted special attention. Brouet et al.~\shortcite{Brouet:2012} formulate a constrained optimization framework to transfer garments across different 3D characters.
Other approaches use data-driven methods~\cite{DRAPE2012,Xu:2014} to replace the expensive physical cloth simulation process and present retargeting examples. In our work, we perform garment retargeting via a novel optimization procedure that directly operates at the joint embedding of different garment design spaces and can be used to transfer across 100s of shape variations while ensuring that desired fold characteristics are preserved. 

\noindent\textbf{Garment capture.} While garment modeling approaches aim to generate realistic garments, capture methods focus on faithfully reconstructing the garment observed in the input data. They accomplish this task often by utilizing more complex capture setups. Pritchard et al.~\shortcite{Pritchard:2003} present one of the earlier approaches where the geometry of a garment is reconstructed from a stereo image pair. Several follow up works have instead used a multi-view capture setup to reconstruct garments with color-coded patterns~\cite{Scholz:2005,White:2007}. Bradley et al.~\shortcite{Bradley:2008} have eliminated the need for a color-coded pattern and presented a multi-view markerless motion capture system for garments. Their follow-up work~\cite{Popa:2009} aims to add details due to wrinkles and folds to the coarse meshes captured from the multi-view video input. Wang et al.~\shortcite{Wang:2010} solve a similar problem of augmenting coarse 3D garment geometry with wrinkles using a data-driven approach. In a separate line of work, given a 2D pattern and 3D contour curve, Rohmer et al.~\shortcite{rohmer:2011} interpolates the curve in a procedural manner to generate folded paper geometry. The more recent approach of Neophytou et al.~\shortcite{Neophytou:2014} analyzes a set of 3D scans to learn a deformation model for human bodies and garments where garments are represented as the residual with respect to the body shape. Finally, the ClothCap~\cite{Pons-Moll:2017} system takes a 3D scan sequence as input and captures both the body and the garment shape assuming weak priors about where a specific type of garment is expected to be with respect to the body. All these methods perform a faithful reconstruction of the garments including the fold and wrinkle geometry but rely on multi-view input or alternate 3D information. In contrast, our method performs a similar faithful prediction using a single sketched image as input and allows for subsequent multi-modal refining in the garment domain and/or the mannequin body shape. 



\begin{figure}[b!]
  \includegraphics[width=\columnwidth]{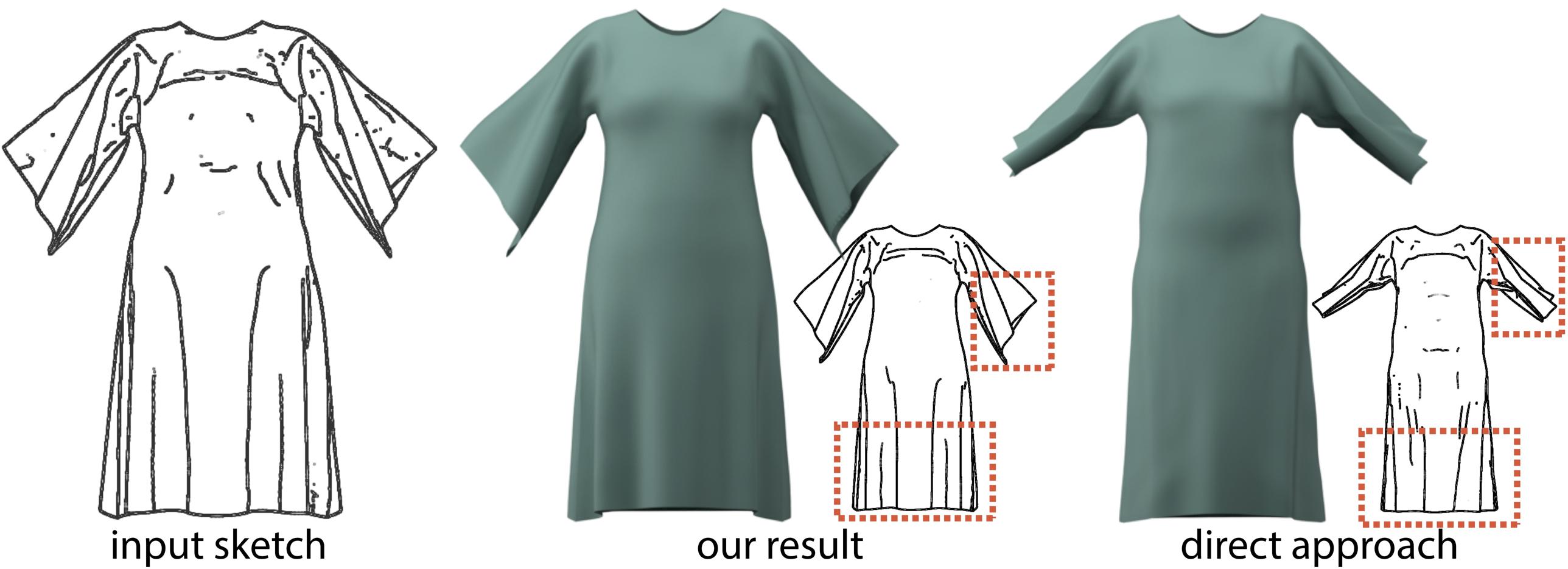}
  \caption{While a network trained directly to infer the draped garment from an input sketch overfits to training data, learning a joint (latent) shape space across different modalities leads to better generalization.}
  \label{fig:direct_approach}
\end{figure}


\section{Approach}
\label{sec:approach}

\subsection{Overview}

Traditional garment design workflows involve interacting with one or more design domains, namely: (a)~the 2D \textit{design sketch} $\mathbb{S}$ that can be used to indicate a desired look and feel of a garment by specifying silhouette and fold lines; 
(b)~the \textit{parameter domain} $\mathbb{P}$ that allows the designer to specify both pattern parameters (i.e., size/material of sewing patterns) and body parameters (i.e., shape of the human body); 
and (c)~the 3D \textit{draped configuration} $\mathbb{M}$ (i.e., the 3D mesh) that captures the final garment shape on the target human body with a garment sized according to its 2D pattern parameters. 

The above-mentioned domains have complementary advantages. For example, sketches provide a natural option to indicate visual characteristics of the folds such as density of folds, silhouette, etc.; parameters are effective to indicate direct changes to garment edits and/or specify target body shapes; while, the draped shape helps to generate previews under varying texture patterns, camera and/or illumination settings. By providing the designer with the ability to indicate target specifications via multiple modalities, we want to exploit the complementary advantages offered by the different domains to enrich the design process.  

The above spaces, however, have very different dimensions making it challenging to robustly transition from one domain to another, or accept inputs across multiple modalities. In other words, a traditional data-driven method can easily overfit to the example datasets and result in unrealistic results in case of new test data. For example, a learned network to help transition from $\mathbb{S} \rightarrow \mathbb{M}$ easily leads to over-fitting as seen on test sketch input in Figure~\ref{fig:direct_approach} (see Section~\ref{sec:results}). More importantly, such an approach does not give the designer access to the pattern and/or body parameters to edit.

Instead, we propose to learn a shared latent space by jointly learning across the three domains using a novel cross-modal loss function (Section~\ref{subsec:learning}). Our key observation is that the shared latent space \textit{regularizes} the learning problem by linking the different domains. From a usage point of view, the artist enters the design space via a sketch, and then continue making further changes by directly editing the inferred pattern and/or body parameters.

%
The designer can now create garments via a multimodal interface by seamlessly indicating sketch behavior, garment or body parameters, or retexturing~(see Figure~\ref{fig:workflow}). 
A garment, thus designed, can then be easily remapped to a range of other body shapes, facilitating pattern grading. To ensure the original sketched folding behaviors do not get lost in the new draped garments adapted for the different body shapes, we present a novel retargeting method that formulates an optimization in the shared latent space (Section~\ref{subsec:retargeting}). Before describing the methods in detail, we next introduce the specific representation choices used in this work.

\begin{figure}[h!]
\centering
    \includegraphics[width=\columnwidth]{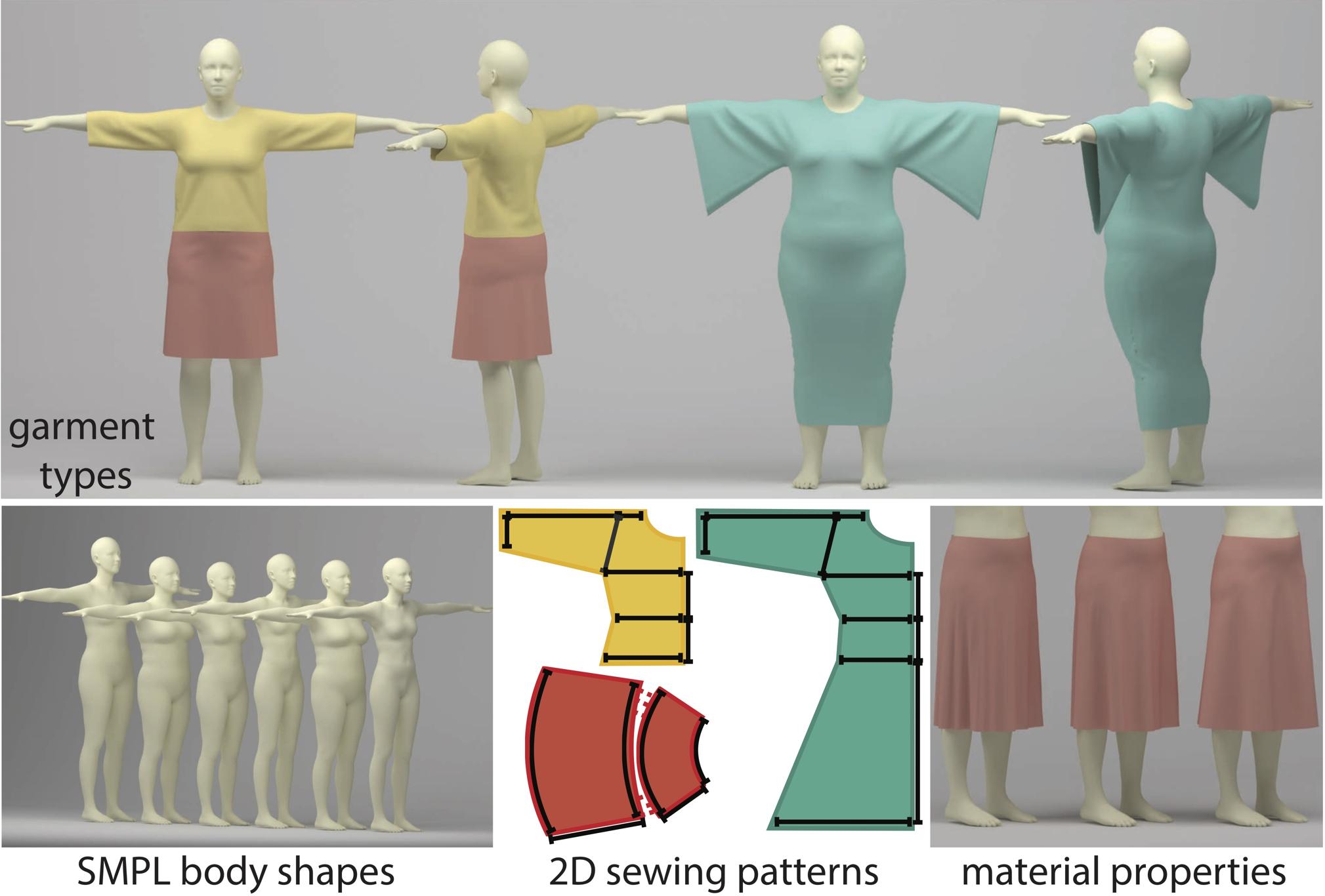}
   \caption{Our dataset includes three types of garments (\texttt{shirt}, \texttt{skirt}, and \texttt{kimono}). Each garment is parameterized by a small number of 2D sewing pattern dimensions as well as material properties including stretch, blend, and shear stiffness parameters. We sample different garment parameters and simulate on different 3D body shapes sampled from the parametric SMPL body model.}
  \label{fig:pattern_parameterization}
\end{figure}

\subsection{Parameter Space}
In this work, we tested on three different garment types namely \texttt{shirts}, \texttt{skirts}, and \texttt{kimonos}. We parameterize each garment type using a small number of key dimensions (e.g., length and width of sleeve for \texttt{shirt}, length of the waistline for \texttt{skirt}) and material properties, i.e., stretch, blend, and shear stiffness (Figure~\ref{fig:pattern_parameterization}). Specifically, the number of parameters for each garment types were: 4 for \texttt{kimono}, 9 for \texttt{shirt}, and 11 for \texttt{skirt}, respectively. We collect all the garment parameters in a vector $\mathbf{G}$. 
We adopt the SMPL~\cite{SMPL:2015} model for the underlying 3D body representation. Given a specific pose, SMPL provides a $10$-dimensional vector $\mathbf{B}$ to describe the body shape variation. Note that the pattern parameters are encoded relative to body size, i.e. vertical parameters are related to body height and horizontal parameters are related to chest circumference or arm length. We denote the combined body and garment parameter space as $\mathbf{P} = (\mathbf{G,B}) \in \mathbb{P}$.

In order to generate a dataset of training examples for a given garment type, we first randomly sample garment parameter instances from $\mathbf{G}$ to generate the 2D patterns. With a target pose, we then sample one of the 2k female body shapes in the FashionPose dataset~\cite{Lassner:UP:2017} to generate samplings of $\mathbb{P}$ resulting in a total of $8000$ combinations. We then simulate combinations of garment samples over body samples using the cloth simulator in FleX~\cite{nvidiaFlex} (see Figure~\ref{fig:simulation_process}).  
This results in draped garment meshes, which we refer to as the mesh $\mathbf{\tilde{M}}$. Given a collection of such meshes~$\{\mathbf{\tilde{M}}\}$ corresponding to the same type of garment, we obtain a compressed representation by performing  Principal Component Analysis (PCA) and represent each garment mesh using the first $k$ ($k=200$ in our test) basis vectors, which we denote as $\mathbf{M}$. Finally, we render each $\mathbf{M}$ from a frontal viewpoint using Suggestive Contours~\cite{DeCarlo:2003:SCF} to approximate the corresponding 2D sketch. We denote the rendered sketch image as $\mathbf{\tilde{S}}$ and apply the convolutional layers of DenseNet~\cite{huang2017densely} to generate a 2208-dimensional descriptor $\mathbf{S}$. Thus, for each instance (parameter combination) in our dataset, we have a 3-element set $\mathbf{(P,M,S)}$. Given this dataset, our goal is to learn a joint latent space shared between the 3 modalities.

\begin{figure}[b!]
  \includegraphics[width=\columnwidth]{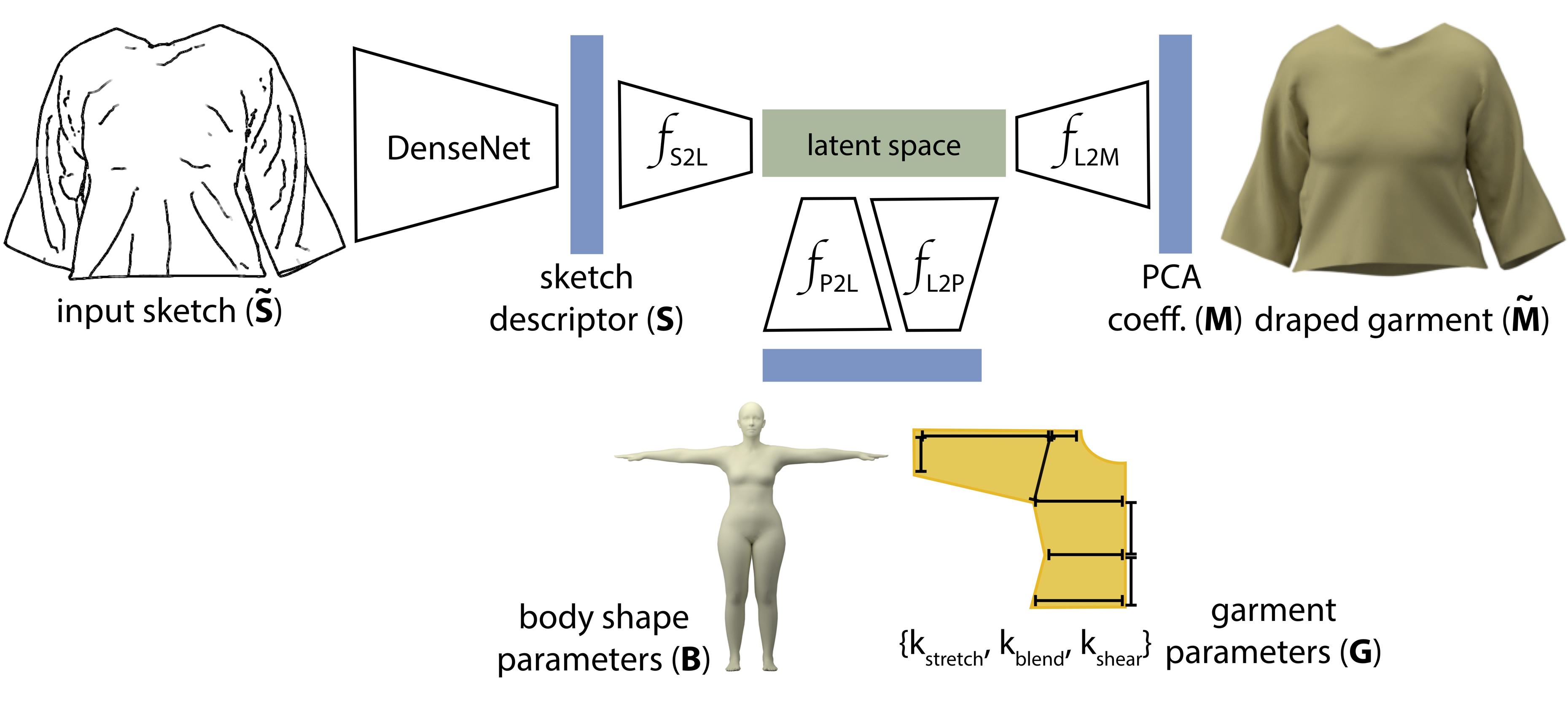}
  \caption{We learn a shared latent shape space between (i)~2D sketches, (ii)~garment and body shape parameters, and (iii)~draped garment shapes by \textit{jointly} training multiple encoder-decoder networks.}
  \label{fig:network}
\end{figure}

\subsection{Joint Latent Space}
\label{subsec:learning}

Given the 3 different modalities $\mathbf{(P,M,S)}$ for each example, our goal is to learn a common $K$-dimensional shared latent space, $\mathbf{L}$, that will enable a multimodal design interface (in our experiments, $K=100$). We achieve this goal by learning the following mapping functions: (i)~sketch descriptor to latent space ($F_{S2L} = \mathbf{S} \rightarrow \mathbf{L}$), (ii)~parameter space to latent space ($F_{P2L} = \mathbf{P} \rightarrow \mathbf{L}$), (iii)~latent space to parameter space ($F_{L2P} = \mathbf{L} \rightarrow \mathbf{P}$), and (iv)~latent space to the draped garment shape ($F_{L2M} = \mathbf{L} \rightarrow \mathbf{M}$). We learn these mappings by jointly training four encoder-decoder neural networks (i.e., sketch-to-parameter, sketch-to-3D garment shape, parameter-to-3D garment shape, and parameter-to-parameter) that share a common embedding space (see Figure~\ref{fig:network}). We describe the network architecture with more details in section \ref{sec:implementation}.

We define a loss function that jointly captures the intention of each of the encoder-decoder networks. Specifically, we penalize (i)~the error in estimating garment and body shape parameters from a sketch, (ii)~the error in estimating the draped garment shape from a sketch or a parameter sample, and (iii)~the reconstruction error of a parameter sample from itself in an auto-encoder fashion. Thus, the combined loss function is defined as:
\begin{equation}\label{eq:1}
\begin{split}
\mathscr{Loss}(\mathbf{P,M,S}) &=  \omega_1\|P-f_{L2P}(f_{S2L}(S))\|_2 
+ \omega_2\|M-f_{L2M}(f_{S2L}(S))\|_2 \\
&+ \omega_3\|M-f_{L2M}(f_{P2L}(P))\|_2 
+ \omega_4\|P-f_{L2P}(f_{P2L}(P))\|_2, \\
\end{split}
\end{equation}
where $\{\omega_1,\omega_2,\omega_3,\omega_4\}$ denote the relative weighting of the individual errors. We set these weights such that the average gradient of each loss is at the same scale (in our experiments $\omega_1=\omega_2=40\omega_3=40\omega_4$). Empirically the consistency terms (the last three terms) make a significant difference on the quality of the network prediction on test data. 


\begin{figure}[h!]
  \includegraphics[width=\columnwidth]{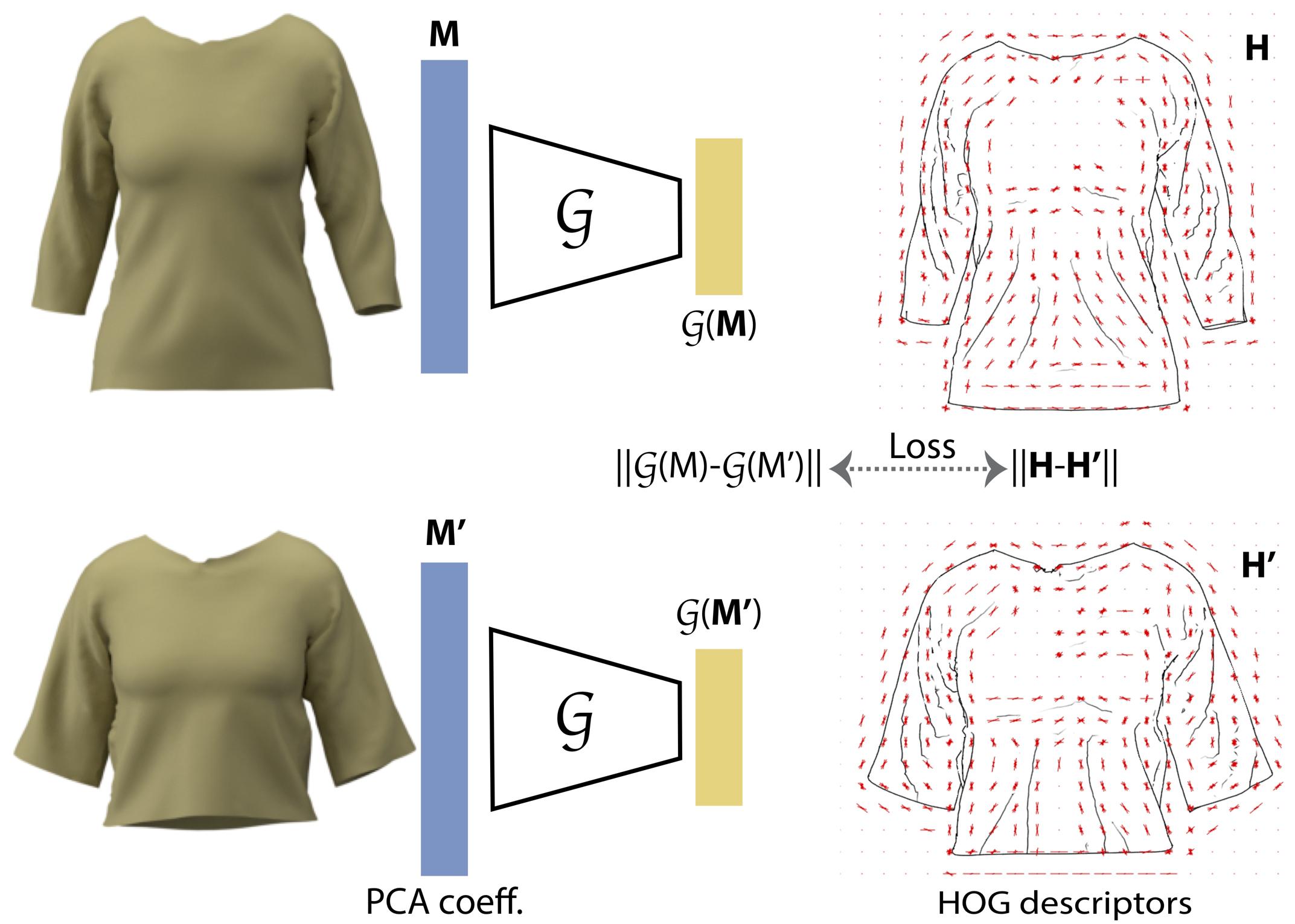}
  \caption{We train a Siamese network $\mathcal{G}$ that learns an embedding of draped garment shapes $\{\mathbf{M}\}$. The distance between a pair of draped garments $(\mathbf{M},\mathbf{M'})$ in this learned embedding space is similar to the distance between the HOG features of the corresponding sketches. Once trained, the loss function can be  differentiated and used for retargeting optimization. }
  \label{fig:siameseNet}
\end{figure}

\subsection{Garment Retargeting}
\label{subsec:retargeting}

One of the most common tasks in real or virtual garment design is retargeting, i.e., adapting a particular garment style to various body shapes. Given a garment $\mathbf{G}$ designed for a particular body shape $\mathbf{B}$, the goal of the retargeting process is to identify a new set of garment parameters $\mathbf{G'}$ for a new body shape $\mathbf{B'}$ such that the look and feel of the draped garments on both body shapes are similar. Naively using the same set of garment parameters on a new body shape does not preserve the original style as shown in Figure~\ref{fig:retargetin_eval}. On the other hand, deforming the draped garment in 3D to match the desired style does not ensure a mapping to a valid configuration of sewing patterns. Instead, we propose a novel optimization process that utilizes the shared latent space presented in the previous section.

As a key component of our optimization, we learn a style-aware distance metric between draped garments. Specifically, given two sets of garment-body instances $(\mathbf{G},\mathbf{B})$ and $(\mathbf{G'},\mathbf{B'})$, our goal is to learn a distance measure between their corresponding draped garments $(\mathbf{M}, \mathbf{M'})$ that is similar to the distance between the sketches of the draped garments, $(\mathbf{S}, \mathbf{S'})$. We achieve this goal by learning an embedding of draped garments, $\mathscr{G(\mathbf{M})}$.

Given pairs of $(\mathbf{M,S})$ and $(\mathbf{M',S'})$, we train a Siamese network such that $\| \mathscr{G(\mathbf{M})} - \mathscr{G(\mathbf{M'})} \|$ is similar to the distance between $(\mathbf{S}, \mathbf{S'})$ (see Figure~\ref{fig:siameseNet}). We measure the distance between two sketches as the distance between their HOG features.


Given the learnt embedding $\mathscr{G}(\mathbf{M})$, we define an objective function for garment retargeting. For a pair of $(\mathbf{B},\mathbf{G})$ and a new body shape $\mathbf{B'}$, we optimize the following energy function:
\begin{equation}
\mathscr{E}(\mathbf{G'|G,B,B'}) =  \|\mathscr{G}(\mathbf{f_{L2M}(f_{P2L}(G,B))})-\mathscr{G}(\mathbf{f_{L2M}(f_{P2L}(G',B'))})\|.
\end{equation}

Since both $\mathscr{G}$ and the mappings $f_{L2M}$, $f_{P2L}$ are learned via differentiable networks, we can efficiently compute the gradient of the objective function during the optimization. In our experiments, we adopt L-BFG-S as the solver.

\if false 

\subsection{Overview}
Given a monocular input, i.e., an image or a sketch, of a garment (or clothing), our goal is to jointly capture the 3D geometry and the 2D UV parameterization of the garment that is consistent with the fold edges observed in the input image. Specifically, \emph{fold edges} occur when part of a cloth occludes another region as the cloth folds. Such edges depict hidden 3D geometry and thus introduce discontinuities in the UV domain (see Figure~\ref{pipeline}). Our goal is both to hallucinate this hidden geometry and preserve the discontinuities in the UV parameterization. Once this information is recovered, it can be used for a variety of applications including texture editing, relighting, and re-simulation.

Garments are composed of developable surfaces and often do not undergo significant stretch. Therefore, the type of typical folds are limited and can be categorized~\cite{bradley2003drawing} motivating us to use a data-driven approach. 
This can be explained by the differential geometry fact that developable surfaces can be locally modeled as planes, cylinders, cones, or generally as tangent surfaces. 
However, exactly matching the whole fold pattern observed in the input to a pre-generated example is extremely difficult if not impossible and requires a massive database of examples. In contrast, we expect to find examples that match local fold patterns in a reasonable dataset that covers common fold types. Hence, in a pre-processing stage, we physically simulate different cloth types (\duygu{what type of cloth}) with varying boundary conditions (e.g., hanging from single or multiple points of support, hanging from an edge or a planar surface). We generate a database of rendered local patches from these simulated examples from different viewpoints (see Section~\ref{subsec:database}). Each such local patch comes with its ground truth set of fold edges as well as the 3D geometry and the corresponding UV-mapping. 
Next, given the input, we first retrieve an overlapping set of candidate local patches from this database that locally matches the input according to a multi-scale similarity measure (see Section~\ref{subsec:retreival}). We utilize fold edges and, whenever available, shading cues as feature descriptor. Once we identify the most similar patches from the database, we perform a novel joint optimization to align the patches both in the 3D geometry space and the 2D UV domain (see Section~\ref{subsec:opt}). Preserving the input fold edges is central to this optimization. 
As a key ingredient to our solution, this joint 3D geometry and 2D UV parameterization registration allows us to obtain high quality geometry hallucination behind the folds and obtain a consistent global UV parameterization. 
Finally, with the recovered geometry and the UV mapping, we can perform various applications (see Section~\ref{sec:results}). We illustrate this pipeline in Figure~\ref{fig:pipeline} and next discuss each stage in detail.

\subsection{Database Generation}
\label{subsec:database}
\duygu{how many different cloth geometries? where do we sample material properties from and how many samples? how do we represent illumination, is it environment map, if so where do we get the environment maps? How do we sample camera locations and how many? how do we choose the boundary conditions and how many different combinations? Which simulator do we use? For each rendered image, what is the patch size?}
For each cloth simulation, we render it with some natural illumination by a virtual camera. we have access to the shading layer and normal layer. Since the UV coordinate of the cloth mesh is known, we can also render the UV correspondence map from the camera view. The discontinuity on the correspondence map forms the seam layer in the image space.

\duygu{how many levels in the image pyramid? what are the patch sizes in each level? what information do we get from each layer to form the descriptor?}
From image space, we randomly select a location and cut 2D patch pyramid with increasing size/decreasing weight from shading layer, normal layer and seam layer. The pyramids then reshape to a vector and will be used as signature during retrieval.

\subsection{Nearest Neighbor Retrieval}
\label{subsec:retreival}

1) overlap a 2d grid on the input. each square with size SxS.

2) at each grid point, we cut a pyramid around that point with small shift on location and different level of rotation.

3) for each grid point, we have boosted it to a group of patches by applying translation and rotation. then for each pyramid, we perform KNN to find $k_1$ closest candidates from the dataset. We then pick the $k_2$ candidates with the smallest knn-distance from all the candidates from all the boosted patches.

4) we use MRF to select the patches. the uniary term is the pyramid distance from candidates to the input while the binary term is the difference of the normal layer of the overlapped neighbor patches.

\subsection{Joint 2D-3D Alignment}

now for each grid point, we have one retrieved patch. they were aligned in image space but not in 3D space and uv space.

we first perform a round of multi-patches ICP to register the patches in 3D (this is tricky, without registration, the final result mesh might have more artifacts. while with registration, the appearance of seam/shading of the final result might be slightly off from the input image.)

UV alignment: jointly update the UV transformation and corresponding points for each patch. (similar to ICP)

\label{subsec:opt}

\fi

\section{Implementation Details}
\label{sec:implementation}
In the following, we provide details about the data generation process and the network architecture.

\begin{figure}[b!]
  \includegraphics[width=\columnwidth]{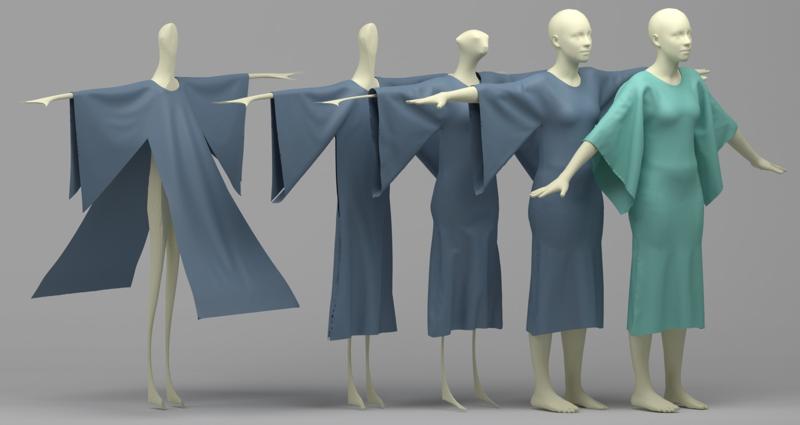}
  \caption{To generate synthetic data, we first shrink the body of the mannequin in rest pose into its skeleton and let the triangle mesh of the 2D sewing patterns drape naturally over it. After the draping converges, we stitch the boundary and inflate the skeleton back to its original shape. We further animate the mannequin into the target body pose to generate our final simulation result.}
  \label{fig:simulation_process}
\end{figure}

\mypara{Data generation. } When generating our training dataset, for a given garment type, i.e., \texttt{shirt}, \texttt{skirt}, or \texttt{kimono}, we first sample a set of garment parameters and generate an isotropic triangle mesh within the silhouette of sewing patterns using centroidal voronoi triangulation. We then simulate each sampled garment on varying body shapes. The cloth simulator we use, i.e., FleX, is particle-based and is sensitive to the edge lengths of the mesh being simulated. Thus, we fix the average edge length across all samples of the garment, leading to meshes with varying topology and face count. In order to ensure a consistent mesh topology to perform PCA on draped garments, we use one of the simulated examples as a reference mesh topology and remesh the remaining examples. Specifically, we locate the vertices of the reference template in its associated triangle in a common sewing pattern space and compute its position in every other simulated example via barycentric coordinates. The simulation of $8000$ samples for each garment type took about $60$ hours to generate.

Once we have a set of simulated, i.e., draped, garments we generate the corresponding sketches using Suggestive Contours~\cite{DeCarlo:2003:SCF} referred as NPR(non-photorealistic rendering) in this paper. We perform data augmentation in the resulting sketches by removing small line segments, curve smoothing, adding gaussian blur, etc. All sketches are centered and cropped into a $224 \times 224$ square patch. We extract a $2208$-dimensional feature vector for each patch via DenseNet~\cite{huang2017densely} (we use the DenseNet-161 architecture provided in the TorchVision library~\cite{torchVision} and use the output of the fourth dense block as our feature vector).

\mypara{Network architecture. } The encoder and decoder networks we train to learn a shared latent space are composed of \emph{linear blocks} which are linear layers followed by Rectifying Linear Unit (RELU) activations and batch normalization. Specifically, the encoder, $F_{S2L}$, takes as input a $2208$-dimensional feature vector of a sketch and maps it to the $K=100$ dimensional shared latent space with $10$ linear blocks (the output dimension size is kept fixed in the first 6 blocks and gradually decreased to $1000$, $500$, $200$, and $100$ in the remaining 4 blocks). The encoder, $F_{P2L}$, takes as input a $p$-dimensional parameter vector representing the garment and the body shape ($p=22$ for \texttt{shirt}, $p=17$ for \texttt{skirt}, $p=24$ for \texttt{kimono} where $3$ material parameters and $10$ body shape parameters are consistent across the different garment types) and maps it to the shared latent space with $6$ linear blocks (the output dimension size is kept fixed in the first block and increased to $100$ in the second block). The decoder, $F_{L2M}$, takes as input the $K=100$ dimensional latent space vector and maps it to the $200$-dimensional PCA basis that represent the draped garment shape. This decoder consists of 6 linear blocks (the output size of the first two blocks is $100$ and the output size of the remaining blocks are $200$). Finally, the decoder, $F_{L2P}$, takes as input the $K=100$ dimensional latent space vector and maps it to the parameters of the garment and the body shape. This decoder consists of 6 linear blocks, where the first 5 blocks keep the output dimension size fixed and the last block changes the output size based on the garment type.

We jointly train the encoder-decoder architectures for $20000$ epochs with a learning rate of $0.1$ and batch size of $64$. We use stochastic gradient descent for network backpropagation.

\mypara{Retargeting. } For retargeting garments across different body shapes, we train a Siamese network that learns an embedding of the draped garments such that the distance between two draped garments is similar to the distance between their corresponding sketches. This Siamese network takes as input a $200$-dimensional PCA vector and maps it to an $100$-dimensional embedding space with $6$ linear blocks (the output dimension size is kept fixed in the first 5 blocks and decreased to $100$ for the last block). We train this network for $20000$ epochs with a learning rate of $0.03$ and batch size of $64$. We use stochastic gradient descent for network backpropagation.

We refer the reader to the supplementary material for a detailed network architecture configuration.

\begin{figure*}[t!]
  \includegraphics[width=\textwidth]{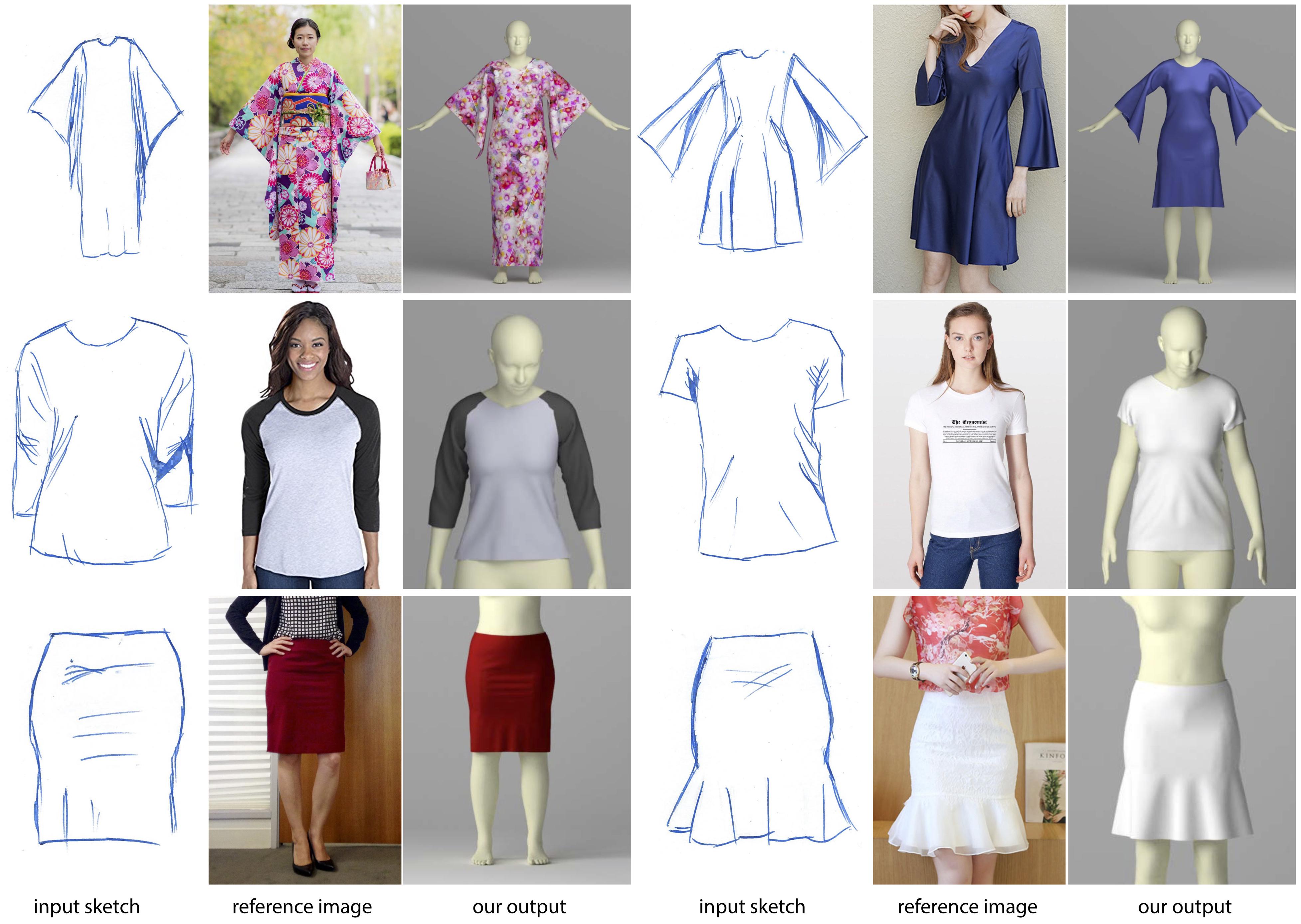}
  \caption{For each reference image, we ask users to draw the corresponding sketch that we provide as input to our method. Our method generates draped garments that closely resemble the reference images.}
  \label{fig:real}
\end{figure*}

\begin{figure*}[t!]
  \includegraphics[width=.975\textwidth]{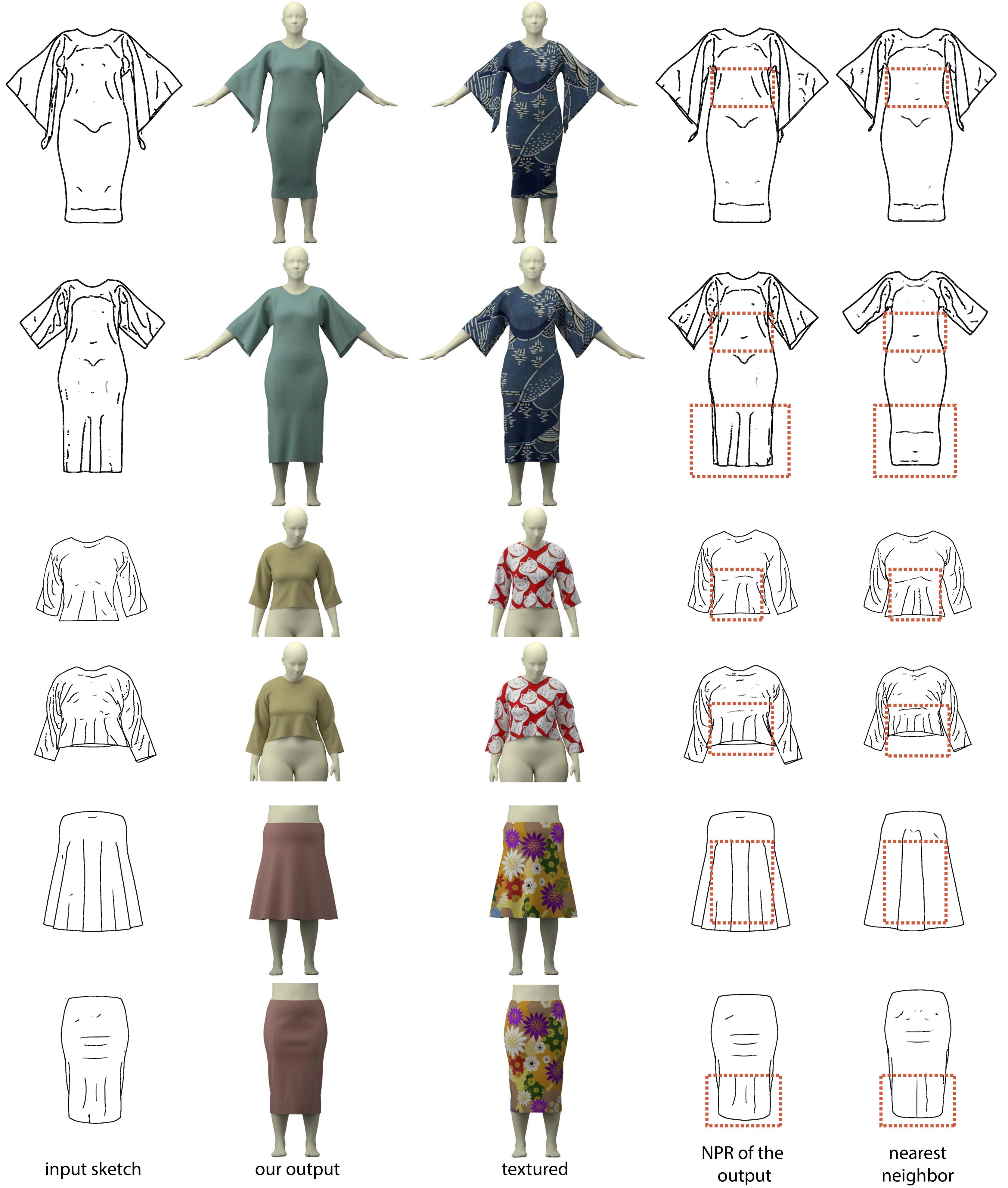}
  \caption{For each sketch, we show our output with/without texture;  NPR rendering of our output and the nearest sketch retrieved from our database. As highlighted in orange, our result plausibly captures the folds provided in the input sketch.}
  \label{fig:results_eval}
  \vspace{-.5cm}
\end{figure*}
\section{Evaluation}
\label{sec:results}

We evaluate our method qualitatively on real and synthetic data and quantitatively on synthetic data. We split our synthetic dataset into training ($95\%$) and testing sets ($5\%$) such that no garment and body parameters are shared across the splits. For evaluations on the synthetic data, we use $200$ samples from the test set for each garment type. We note that given an input sketch, our method estimates the corresponding garment and body shape parameters as well as the shape of the draped garment. In a typical design workflow, once satisfied with the design, we expect the designer to perform a cloth simulation using the estimated garment and body shape parameters to obtain the final and accurate look of the garment. We note that the draped garment shape predicted by our network closely resembles the simulation result performed by using the predictions of the garment and body parameters (see Figure~\ref{fig:net-sim}). Thus, we only provide the final simulated results.
In Figure~\ref{fig:results_eval}, for each input synthetic sketch, we show the simulated results with the garment and body shape parameters predicted by our method. Since our network can predict the corresponding 2D sewing pattern parameters as well, we can easily generate uv coordinates for the draped garment and texture it. To show that our network does not learn to memorize the training examples, we also show the nearest neighbors retreived from our training dataset using the DenseNet features of sketches. As highlighted in the figure, while the NPR renderings of our results closely resemble the input sketch, the nearest neighbors fail to capture many folds present in the input sketch.

\begin{figure*}[t!]
  \includegraphics[width=\textwidth]{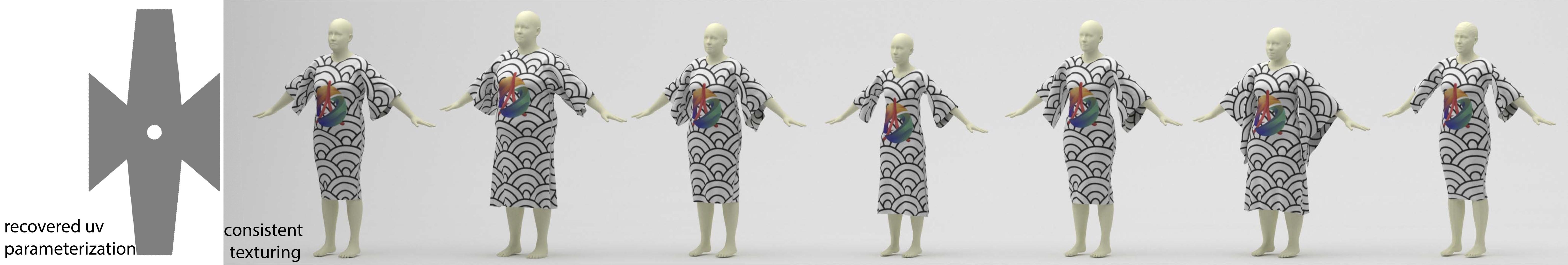}
  \caption{Our method generates consistent uv parameters across different instances of the same garment type. In this example, the alignment between the texture patterns can be best seen in the neck region.}
  \label{fig:parameterization}
\end{figure*}
\begin{table*}[t!]
\begin{center}
\caption{Starting with an input sketch (top) or a set of garment and body shape parameters (bottom), we report the average L2 error in the estimated draped garment (both in terms of PCA basis and  vertex positions), body shape (both in terms of SMPL parameters and  vertex positions), and the garment parameters. All the parameters are normalized to the range $[0,1]$.}
\label{tab:quantitativeNumbers_2}
\begin{tabular}{ |p{2.5cm}|c|c|c|c|c|c|c|c|c|c|c|c|  }
\hline
From \textbf{sketch} & \multicolumn{4}{|c|}{Garment mesh} & \multicolumn{4}{|c|}{Body shape} & \multicolumn{2}{|c|}{Garment parameter} \\
\hline
& \multicolumn{2}{|c|}{Training set} & \multicolumn{2}{|c|}{Testing set} & \multicolumn{2}{|c|}{Training set} & \multicolumn{2}{|c|}{Testing set} & \multicolumn{1}{|c|}{Training set} & \multicolumn{1}{|c|}{Testing set}\\
\hline
& L2 PCA & L2 Vert. & L2 PCA & L2 Vert & L2 SMPL & L2 Vert.& L2 SMPL & L2 Vert & L2 & L2 \\
\hline
Shirt & 1.99\% & 1.58\% & 4.21\% & 3.82\% & 2.02\% & 1.77\% & 5.22\% & 4.67\% & 5.26\% & 6.73\% \\
\hline
Skirt & 1.76\% & 1.14\% & 2.38\% & 2.11\% & 1.59\% & 1.13\% & 3.29\% & 2.06\% & 3.79\% & 4.99\% \\
\hline
Kimono & 2.28\% & 1.70\% & 5.45\% & 4.48\% & 3.84\% & 2.74\% & 7.48\% & 5.35\% & 6.94\% & 8.47\% \\
\hline
\end{tabular}
\end{center}
\begin{center}
\begin{tabular}{ |p{2.5cm}|c|c|c|c|c|c|c|c|c|c|c|c|  }
\hline
From \textbf{parameter} & \multicolumn{4}{|c|}{Garment mesh} & \multicolumn{4}{|c|}{Body shape} & \multicolumn{2}{|c|}{Garment parameter} \\
\hline
& \multicolumn{2}{|c|}{Training set} & \multicolumn{2}{|c|}{Testing set} & \multicolumn{2}{|c|}{Training set} & \multicolumn{2}{|c|}{Testing set} & \multicolumn{1}{|c|}{Training set} & \multicolumn{1}{|c|}{Testing set}\\
\hline
& L2 PCA & L2 Vert.& L2 PCA & L2 Vert & L2 SMPL & L2 Vert.& L2 SMPL & L2 Vert & L2 & L2 \\
\hline
Shirt & 1.56\% & 1.16\% & 3.57\% & 3.01\% & 1.47\% & 1.20\% & 3.51\% & 3.29\% & 3.88\% & 3.89\% \\
\hline
Skirt & 1.29\% & 0.98\% & 2.03\% & 1.61\% & 1.09\% & 0.86\% & 2.14\% & 1.68\% & 2.47\% & 3.60\% \\
\hline
Kimono & 1.80\% & 1.58\% & 5.21\% & 2.59\% & 2.80\% & 2.01\% & 4.81\% & 3.33\% & 4.13\% & 4.33\% \\
\hline
\end{tabular}
\end{center}
\end{table*}

\begin{figure}[b!]
  \includegraphics[width=\columnwidth]{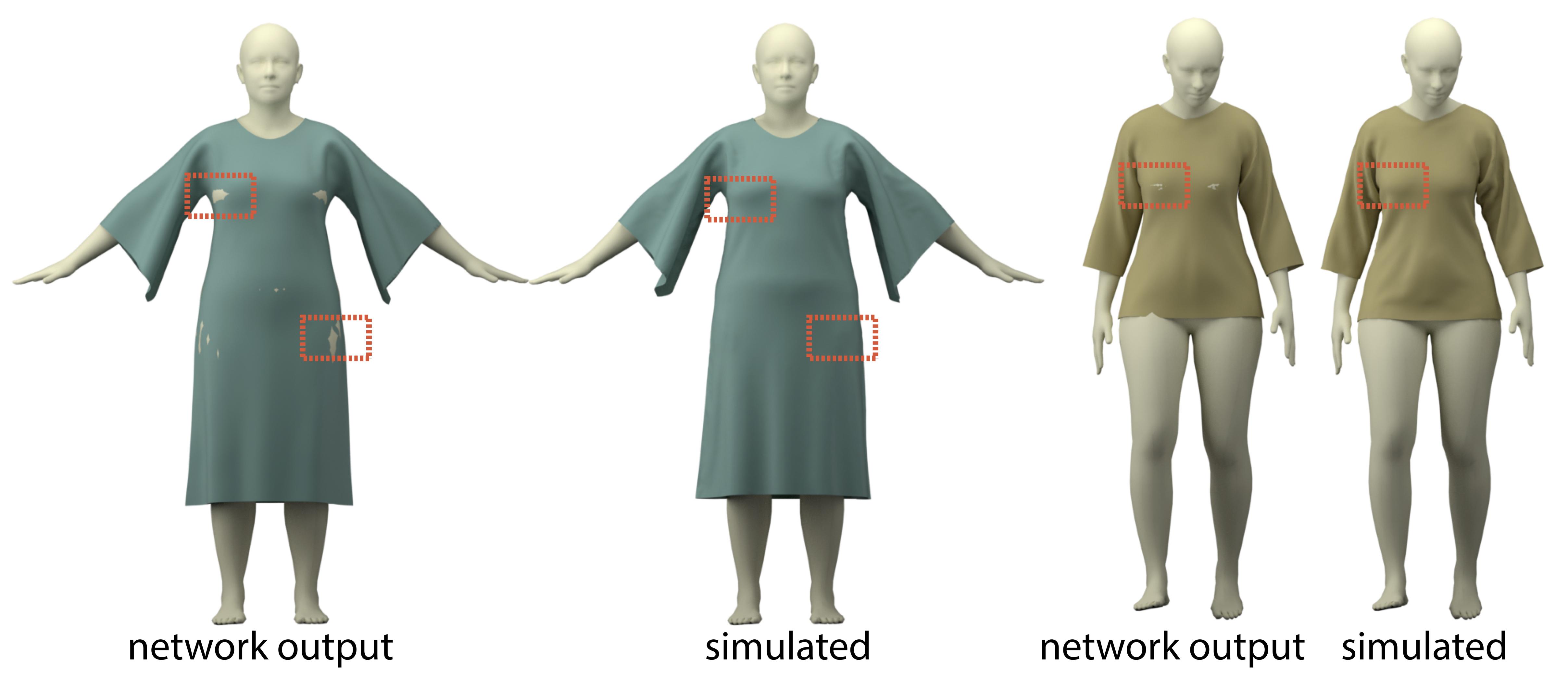}
  \caption{Draped garment shape predicted by our network closely resembles the simulation result using predicted garment/body parameters, and hence can be used to optionally fix draped garment and body intersections.}
  \label{fig:net-sim}
\end{figure}

In Figure~\ref{fig:real}, we provide examples on real images to test the generality of our approach. For 6 different reference images, we ask different users to draw the corresponding sketch that is provided as input to our method. As shown in the figure, our method is able to generate the draped garments that closely follow the reference image and the input sketches. 

We demonstrate the consistency of the uv parameters generated by our method by texturing the same garment type with varying parameters draped on varying body shapes (see Figure~\ref{fig:parameterization}). Our method generates uv parameters that are free of distortions compared to generic solutions.

\mypara{Quantitative evaluations. }
We quantitatively evaluate the performance of the different mappings learned by our network. Specifically, starting either with an input sketch (or a set of garment and body shape parameters), we first map the input to the latent space via $f_{S2L}$ (or $f_{P2L}$). Then, we measure the error in the predicted draped garment mesh (the output of $f_{L2M}$) and the estimated garment and body shape parameters (the output of $f_{L2P}$). For the draped garment, we report the average L2 error both on the PCA coefficients and the vertex positions, similarly, for the body shape, we report the average L2 error both on the SMPL shape parameters and the vertex positions. We normalize all the parameters and the garment and body shapes to the range $[0,1]$ and report the percentage errors. Table~\ref{tab:quantitativeNumbers_2} provides results when a sketch or a set of garment and body parameters are provided as input.

\begin{figure*}[t!]
  \includegraphics[width=\textwidth]{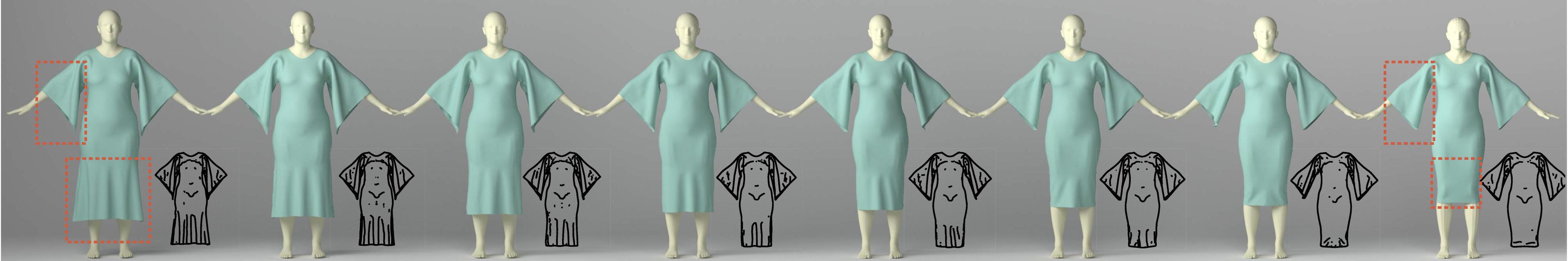}
  \caption{Given two kimono designs shown on the left and the right, we interpolate between their corresponding descriptors in the shared latent space. For each interpolation sample, we show the 3D draped garments inferred by our network and the NPR rendering of the simulated mesh from the corresponding sewing parameters. Our shared latent space is continuously resulting in smooth in-between results.}
  \label{fig:kimono_results_interp}
\end{figure*}

\mypara{Joint latent space evaluation. } In order to demonstrate the benefit of jointly training a shared latent space across three modalities, we train an alternative single encoder-decoder network composed of the mappings $f_{S2L}$ and $f_{L2M}$. As shown in Figure~\ref{fig:direct_approach}, this direct mapping overfits to the training data and is not able to generalize. In contrast, jointly training for additional mappings regularizes the problem and leads to  better performance during test time.

Our learned shared latent space is compact and smooth, as shown in Figure\ref{fig:kimono_results_interp}. When we linearly interpolate between two samples in the latent space, we obtain intermediate results that change smoothly both in terms of draped garment shapes  The internal positions indicate a smooth change in both 3D predict mesh and garment sewing parameters.

\begin{figure}[h!]
  \includegraphics[width=\columnwidth]{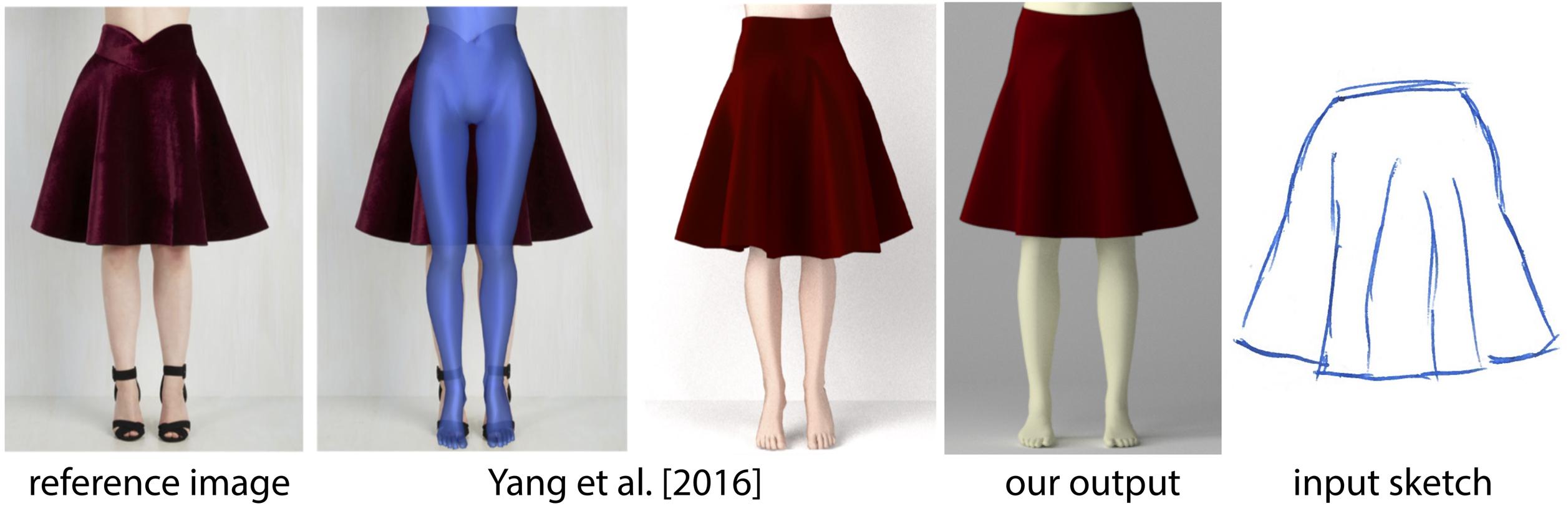}
  \caption{We test our method on one of the examples provided in Figure 8 of Yang et al.~\shortcite{Yang:2016}. We get results that are similar quality at interactive rates compared to their computationally-heavy approach.}
  \label{fig:comparison}
\end{figure}

\mypara{Comparison. }
In Figure~\ref{fig:comparison}, we compare our approach with one example from Figure 8 of Yang et al.~\shortcite{Yang:2016}. Given the reference image in this example, we ask a user to provide the corresponding sketch that is provided as input to out method. The estimate draped garment shape by our method is of similar quality but is generated at interactive rates in contrast to the computation-heavy approach of Yang et al.~\shortcite{Yang:2016}.

%
%
%

\begin{figure}[h!]
  \includegraphics[width=\columnwidth]{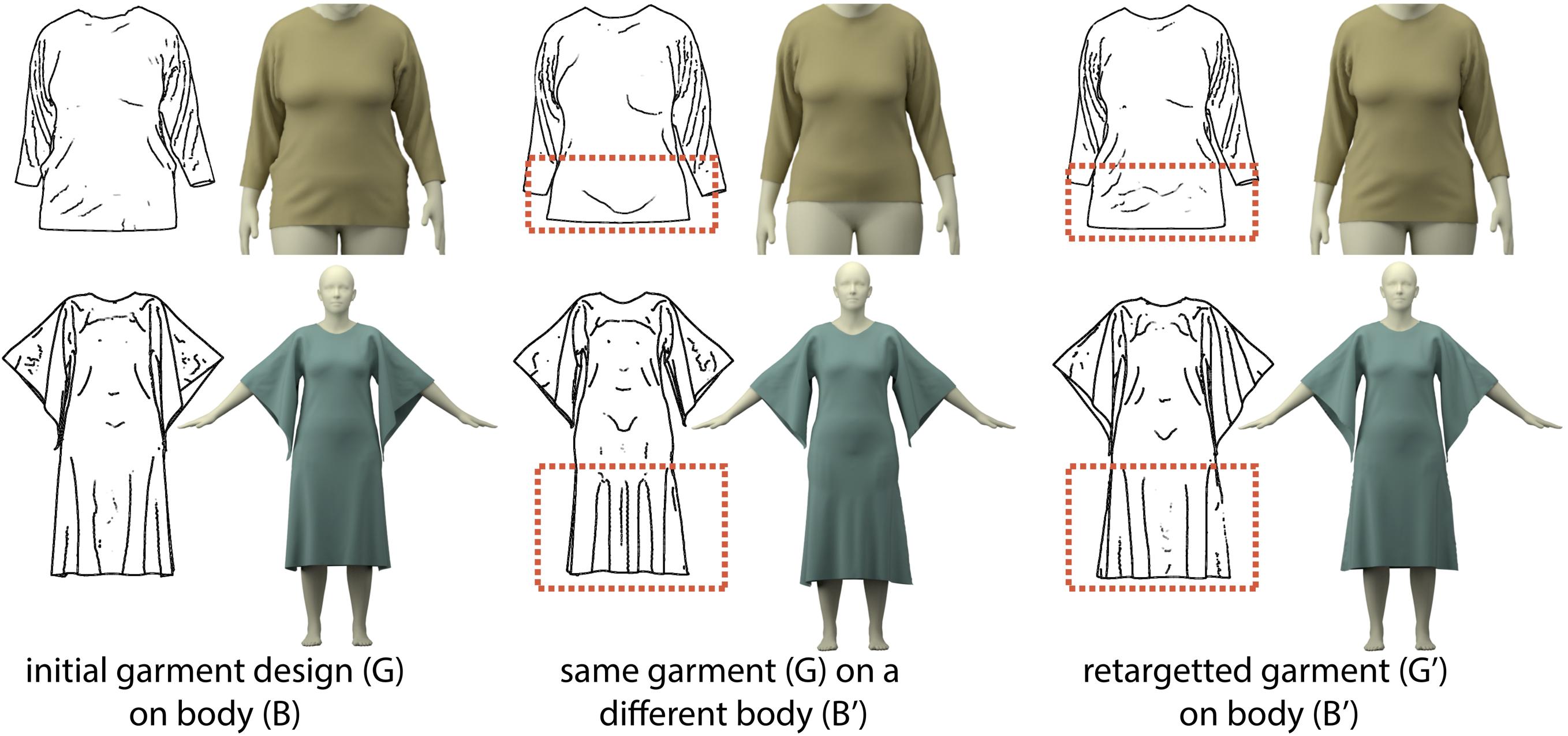}
  \caption{Given an initial garment design, draping the same garment on a different body shape does not preserve the style of the design. In contrast, our retargeting method optimizes for a new set of garment parameters that result in the same look of the garment on a new body shape.}
  \label{fig:retargetin_eval}
\end{figure}

\mypara{Retargeting evaluation.} As shown in Figure~\ref{fig:retargetin_eval}, draping the same garment on different body shapes do not preserve the style of the garment. Our retargeting optimization, in contrast, identifies a new set of garment parameters that would preserve the style of the original garment on a body shape. 

%

\begin{figure}[t!]
  \includegraphics[width=.9\columnwidth]{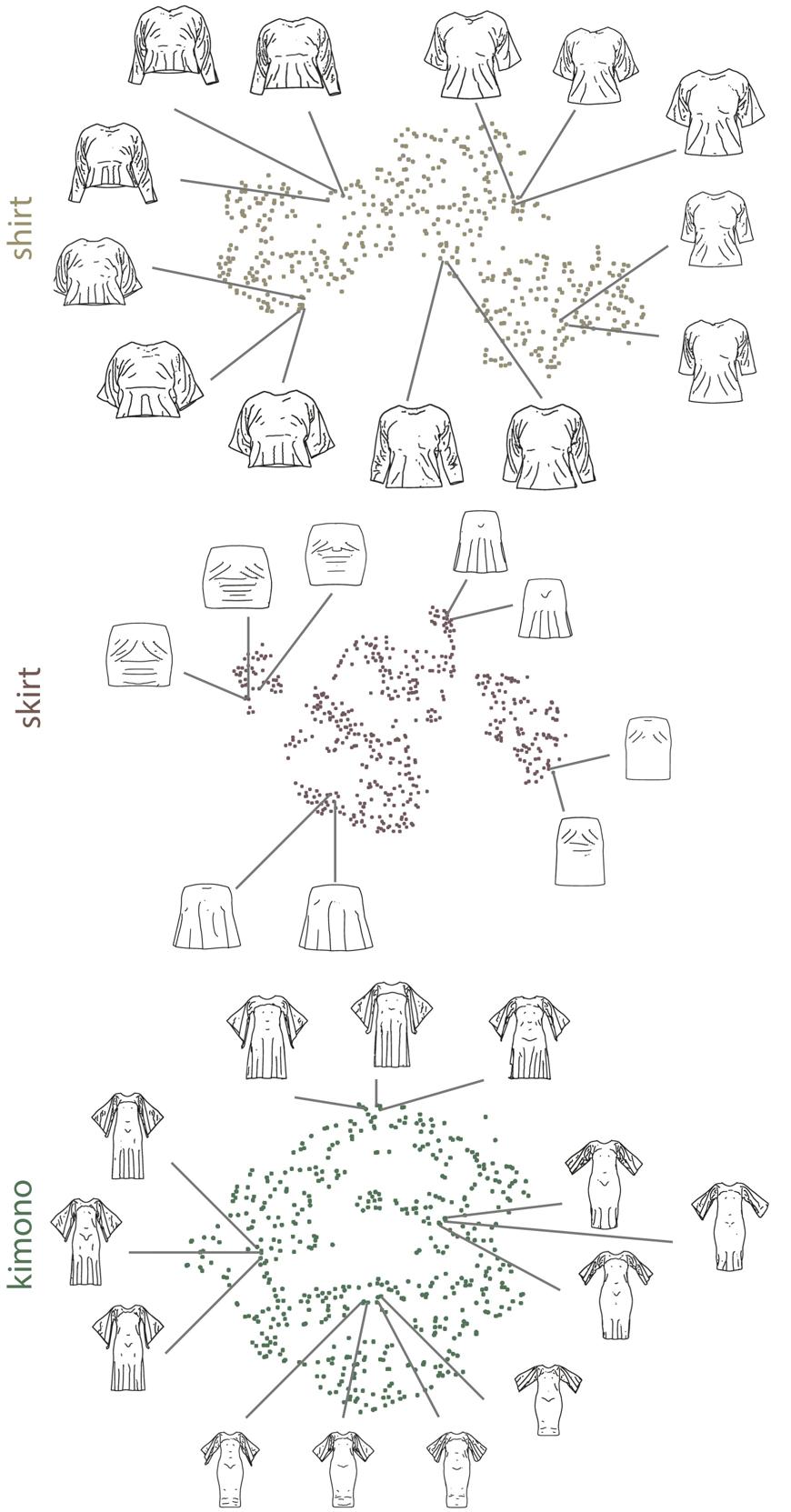}
  \caption{The distance embedding we learn in Section~\ref{subsec:retargeting} clusters similar style garments together.}
  \label{fig:tsne}
  \vspace{-.1in}
\end{figure}

Our retargeting optimization uses a distance measure between draped garments based on the Siamese network introduced in Section~\ref{subsec:retargeting}. We show a tSNE visualization~\cite{tsne} of the embedding learned by this network in Figure~\ref{fig:tsne}. For each embedding, we provide the sketches corresponding to various random samples showing that similar sketches get clustered.

\begin{figure*}[h!]
  \includegraphics[width=\textwidth]{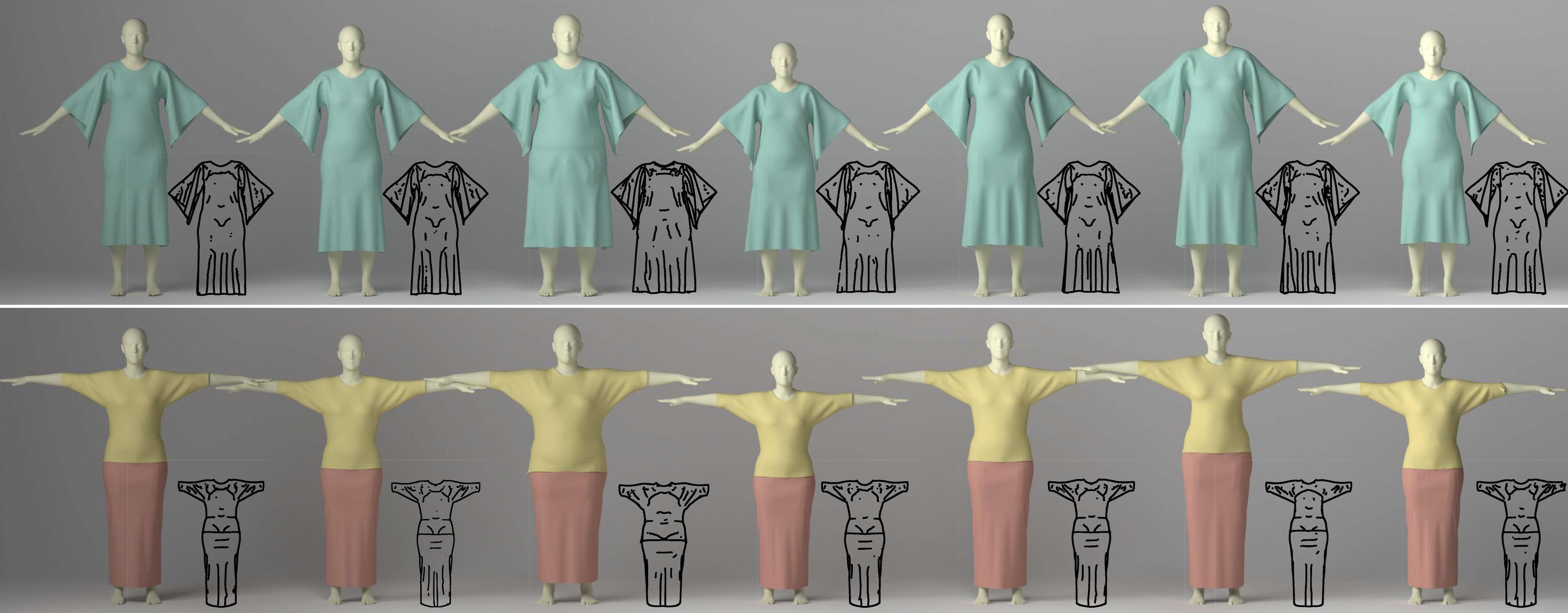}
  \caption{Given a reference garment style on the left, we retarget it to various body shapes. For each example, we also provide the NPR renderings of the draped garments.}
  \label{fig:retargeting_results}
\end{figure*}

\begin{figure}[h!]
  \includegraphics[width=\columnwidth]{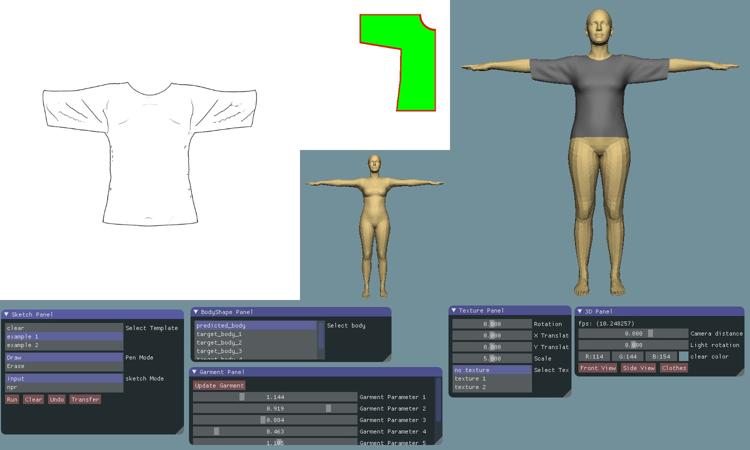}
  \caption{Our user interface allows the user to edit the sketch, garment or body shape parameters, or the texture coordinates and visualizes the updated draped garment in 3D at interactive rates enabling multimodal garment design.}
  \label{fig:ui}
\end{figure}

\mypara{Interactive user interface. }
Based on the shared latent space we learn across different design modalities, we implement an interactive user interface (see Figure~\ref{fig:ui}). In this interface, the user can interactively edit the input sketch, the garment or the body shape parameters, or the uv coordinates and interactively visualize the corresponding draped garment. We refer to the supplementary video which shows each of these editing options.

\begin{figure}[h!]
  \includegraphics[width=\columnwidth]{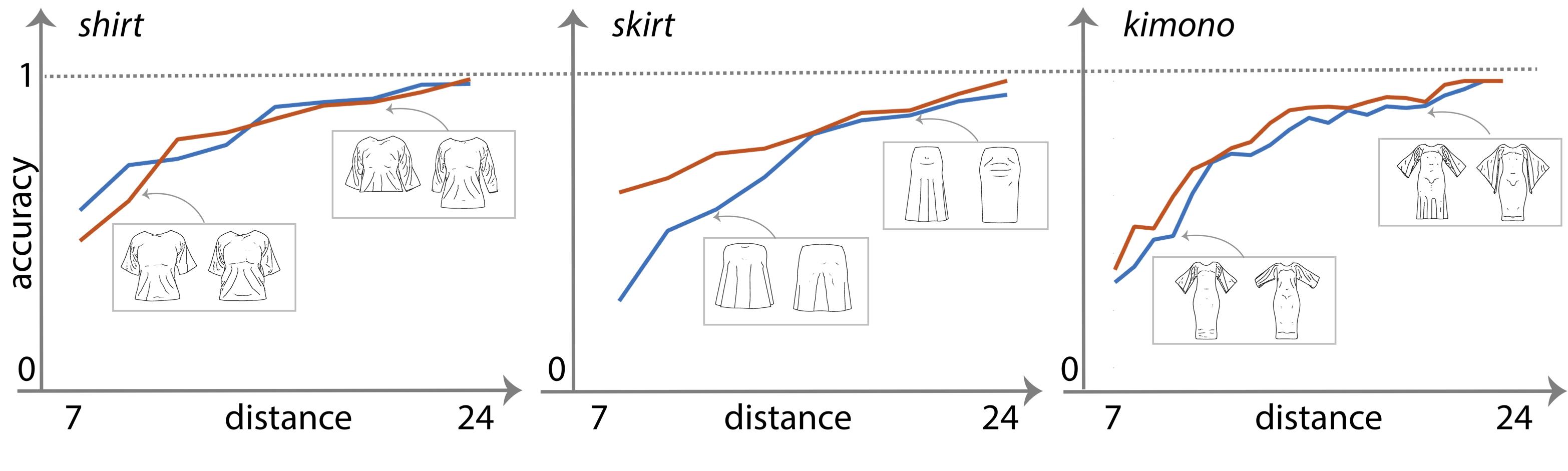}
  \caption{We asked users to pair two given sketches with two draped garment shapes both in the form of the network output (in blue) and the simulation result (in pink). We plot the accuracy of the pairings versus the similarity of the given sketches.  The users are accurate with pairings unless the input sketches are very similar.}
  \label{fig:user_study}
\end{figure}

\mypara{User study.}
Finally, we conduct a user study to evaluate how closely our results resemble the input sketches. Specifically, given two input sketches $(S_i,S_j)$, we show the corresponding draped garment shapes $(M_i,M_j)$ predicted by our network and ask the user to pair the sketches to the garments (e.g., $S_i$ should be paired with $M_i$). If the network output successfully captures the folds in the input sketch, we expect the accuracy of such pairings to be high. However, if $S_i$ and $S_j$ are similar to begin with, the pairing becomes ambiguous. We generate $400$ pairing queries for each garment type and ask $13$ Amazon Mechanical Turk users to answer each query. We plot the accuracy of the pairings vs. the similarity of the input sketches (in terms of the L2 distance between DenseNet features) in Figure~\ref{fig:user_study}. Since we expect the simulation result (performed using the parameters predicted by our network) to capture more details, we repeat the same user study using the simulation result as the draped garment shape. As shown in the plots, the users are accurate unless the input sketches are very similar. The accuracy slightly increases when simulation results are used validating that the generated draped garments perceptually capture the fold characteristics in the input sketches.

\if false  

\begin{enumerate}
  \item Dataset
  \begin{enumerate}
  \item shirt in rest pose
  \item skirt in rest pose
  \item shirt in a different pose
  \item kimono?
  \item real sketch (user study)
  \item \tuanfeng{sketch from image edge detection}
  \end{enumerate}
  
  \item Reconstruction
  \begin{enumerate}
  \item sketch to shape
  \item sketch to garment/body parameter
  \item parameters to shape
  \item parameter to parameter
  \item \tuanfeng{sketch to parameter to shape}
  \item show nearest neighbour
  \end{enumerate}
  
  \item Interpolation
  \begin{enumerate}
  \item in sketch (feature) space
  \item in parameter space
  \item in latent space (from sketch)
  \item in latent space (from parameter)
  \end{enumerate}
  
  \item UV
  \begin{enumerate}
  \item distortion compare to gt and abf
  \item visualization
  \end{enumerate}
  
  \item Refinement with simulator
  
  \item Network architecture/Parameters
  \begin{enumerate}
  \item PCA dimension
  \item latent space dimension
  \item Loss function (weight?)
  \end{enumerate}
  
  \item Retargeting
  \begin{enumerate}
  \item embedding network
  \item optimization (energy decrease)
  \item via simulator
  \item via network
  \item different feature
  \item discuss limitation?
  \end{enumerate}
  
  \item Arcsim toy dataset

\end{enumerate}

\fi
\section{Conclusion}

We presented a data-driven learning framework for obtaining a joint latent space linking 2D sketches, garment (2D sewing patterns and material properties) and human body specifications, and 
3D draped garment shapes in the context of garment design. 
The learned latent space enables a novel multimodal design paradigm that allows users to iteratively create complex draped garments 
without requiring expensive physical simulations at design time. 
We also utilize the latent space to formulate an optimization that allows designed garments to be retargeted to a range of different human body shapes while preserving the 
original design intent. 
Finally, we evaluated our method, both quantitatively and qualitatively, in different usage scenarios and showed compelling results.

\begin{wrapfigure}{r}{0.2\textwidth} 
    \centering
    \includegraphics[width=0.2\textwidth]{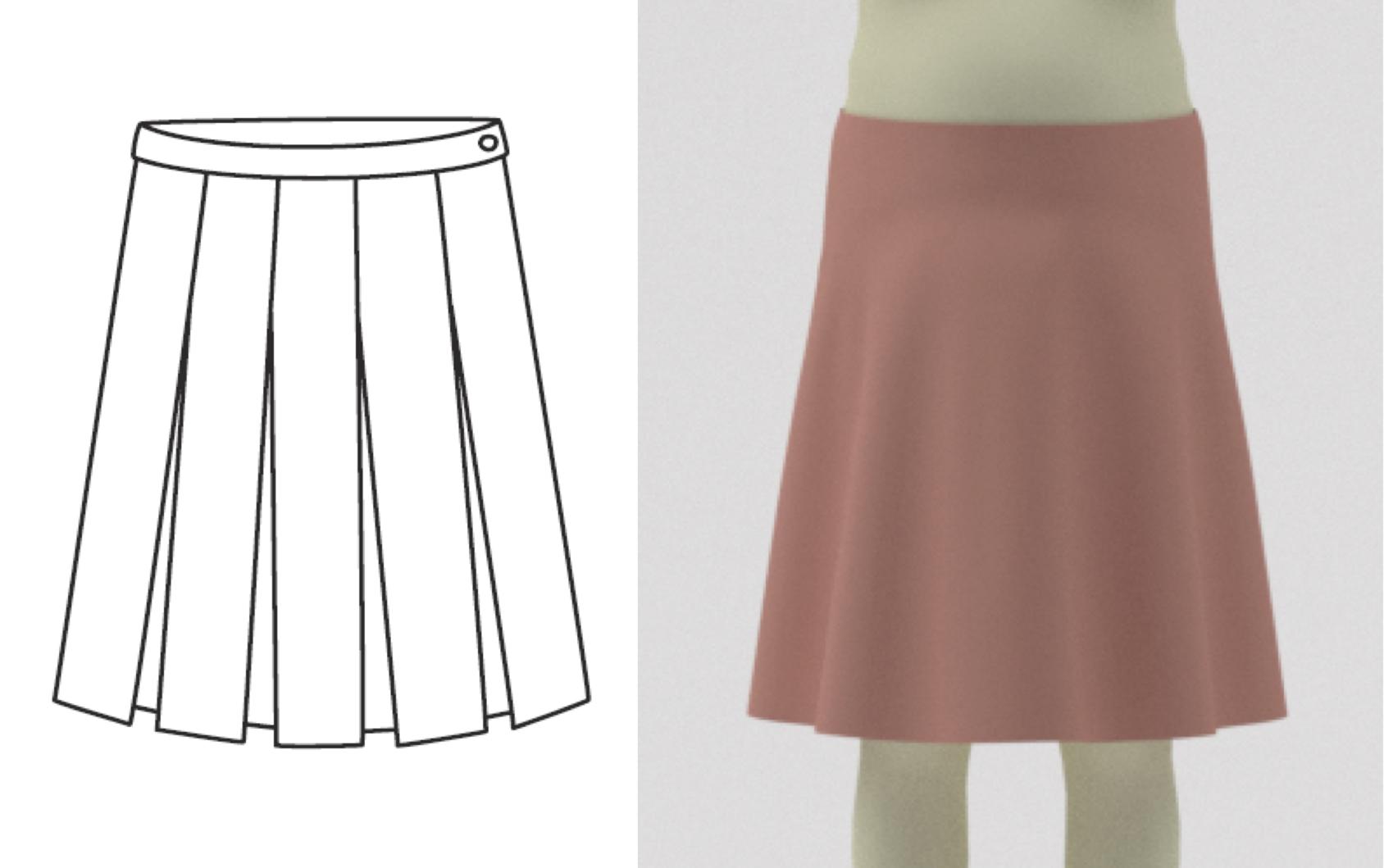}
    \hspace{-10pt}
    \vspace{-10pt}
\end{wrapfigure}
\begin{figure}[h!]
    \centering
    \includegraphics[width=.45\textwidth]{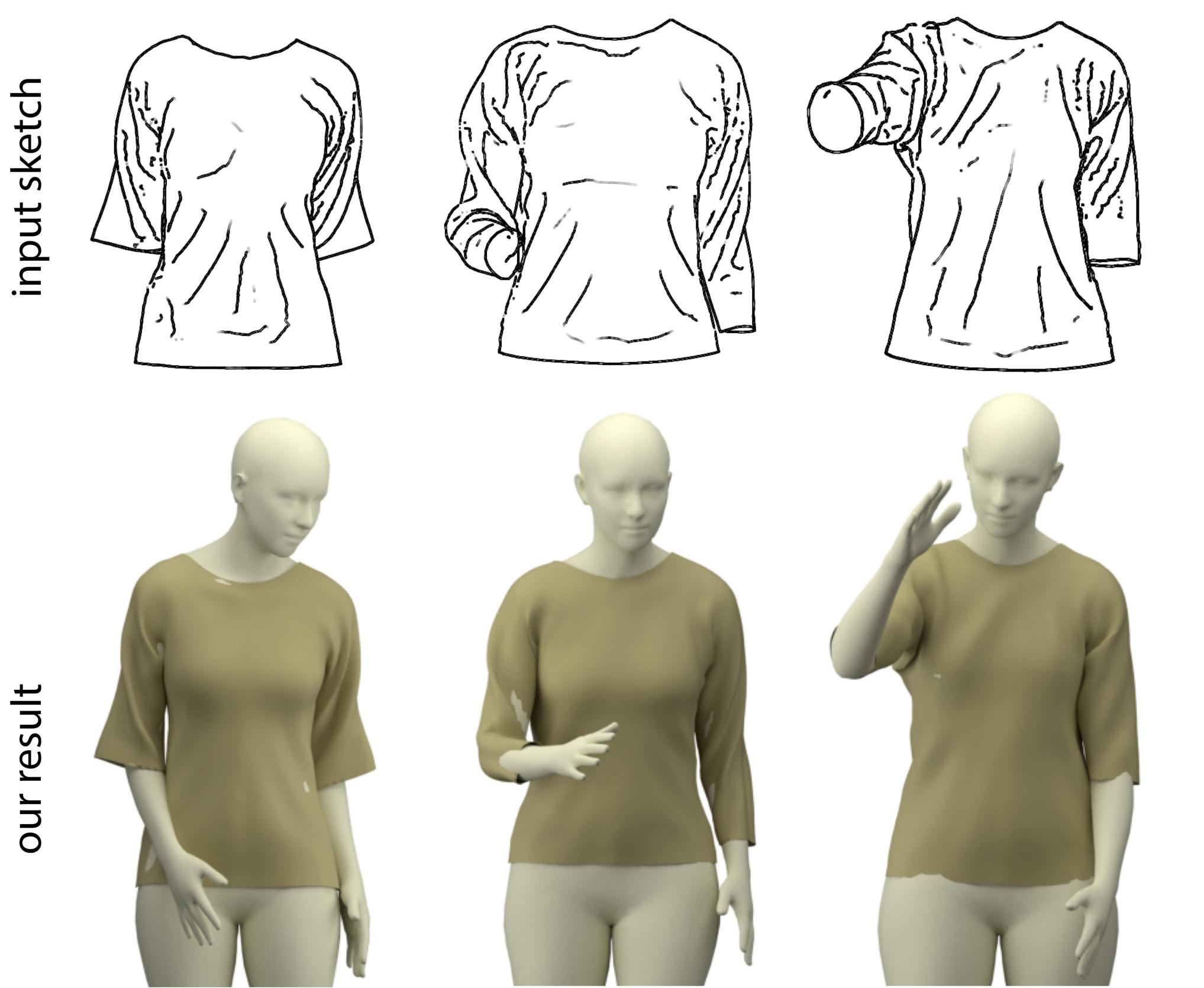}
    \caption{By augmenting our training dataset, we show preliminary results our method inferring body pose.}
    \label{fig:future_pose_variations}
\end{figure}

\mypara{Limitations and future work. }Our method has certain limitations that open up interesting future directions. For each garment type, we have a pre-defined set of 2D sewing parameters. Thus, we cannot represent garment shapes that are not covered by this set as in the inset example. Currently, our method does not handle pose variation, we assume the garments are designed for a body at a fixed target pose. In the future, we plan to expand the latent space by also regressing across common body pose variations. In Figure~\ref{fig:future_pose_variations}, we show preliminary results in this direction. We augment our dataset by considering an additional degree of freedom for body pose that interpolates between different arm configurations. Then, given  the input sketches in the top row, we infer the draped garments and body shape and pose parameters shown in the bottom row. These results indicate that by augmenting the dataset, learning pose variations is possible. The challenge, however, is how to decide which key poses to include in order to effectively sample the configuration space.

In this work, we learned latent spaces specialized to different garment types  (\texttt{shirt}, \texttt{skirt}, and \texttt{kimono}) --- a difficult research question is how to unify these different garment-specific latent spaces via a common space. This is challenging given the complex discrete and continuous changes necessary to transition from one garment type to another. One possibility is to perform interactive exploration where the user annotates additional cut/fold lines as recently demonstrated in \cite{Li:2018}.  

Finally, we would like to explore the effectiveness of the joint learning in other design contexts involving multimodal inputs.

\bibliographystyle{ACM-Reference-Format}
\bibliography{garmentCapture}
\end{document}